\documentclass[pra,twocolumn,showpacs,preprintnumbers,amsmath,amssymb]{revtex4-1}
\usepackage{multirow}
\usepackage{graphicx}
\usepackage{dcolumn}
\usepackage{bm}
\usepackage{psfrag}
\usepackage{epsfig}
\usepackage{amsmath}
\usepackage{amssymb}
\usepackage{MnSymbol}
\usepackage{color}
\usepackage{bbm}
\usepackage[FIGTOPCAP,raggedright,nooneline]{subfigure}
\usepackage{verbatim}
 \usepackage[applemac]{inputenc}
\usepackage{textcmds} 
\usepackage{tikz}
\usepackage{natbib}

\newcommand{\ignore}[1]{}

\begin{document}

\title{Emergence of solitons from many-body photon bound states in quantum nonlinear media}

\author{G. Calaj\'o$^1$ and   D. E. Chang$^{1,2}$}

\affiliation{$^1$ICFO-Institut de Ciencies Fotoniques, The Barcelona Institute of
Science and Technology, 08860 Castelldefels (Barcelona), Spain}
\affiliation{$^2$ICREA-Instituci\'{o}  Catalana  de  Recerca  i  Estudis  Avan\c{c}ats,  08015  Barcelona,  Spain}

\date{\today}
 
\begin{abstract}
Solitons are  known to occur in the context of atom-light interaction via the well-known semi-classical  phenomenon of self-induced transparency (SIT). 
Separately, in the regime where both light and atoms are fully treated quantum mechanically, quantum few-photon bound states are known to be a ubiquitous phenomenon that arises in different systems such as atoms coupled to chiral or bidirectional waveguides, and in Rydberg atomic media. In the specific case of two-level atoms coupled to a chiral waveguide, a recent analysis based on Bethe ansatz has established that SIT emerges from the quantum realm as a superposition of quantum many-photon bound states.
Beyond this case, however, the nature of any connection between the full quantum many-body regime and semi-classical behavior has not been established. Here, we employ a general spin-model formulation of quantum atom-light interfaces to numerically investigate this problem, taking advantage of the fact that this approach readily allows for powerful many-body simulations based on matrix product states (MPS). 
 We first analytically derive the two-photon bound state dispersion relation for a variety of atom-light interfaces, and then proceed to  numerically investigate the  multi-excitation bound states dynamics. 
Interestingly, for all the specific systems studied, we find that the large-photon number limit always coincides with  the  soliton phenomenon of self-induced transparency or immediate generalizations thereof.



\end{abstract}

\maketitle

\section{Introduction}

One of the most distinct predictions within classical nonlinear optics is the emergence of solitons, whose shape does not change during propagation \cite{Agrawal}. From early on, there were also efforts to understand how these solitons might emerge from a fully quantum theory, perhaps most prominently in continuous Kerr nonlinear media~\cite{lai1,lai2,Kurizki1,Kurizki2,Kurizki3}. While the weak nonlinearities of conventional media historically made quantum properties largely inaccessible, in recent years there have been diverse systems, ranging from ensembles of Rydberg atoms~\cite{firstrev,callum,first,LiangBS}  to waveguide QED systems coupled to quantum emitters~\cite{roy_rev_mod,review_lodahl,Darrick_rev_mod,shen_fan_prl,shen_fan,
Harold_strcorr,mahmo_calajo,Harold_bs,Prasadnat}, where strong interactions at the level of individual photons~\cite{changnl} can be achieved. 
An intriguing effect  occurring in such quantum nonlinear media is the existence of  quantum photon bound states, which have been 
predicted and studied in these different systems with equally diverse theoretical techniques, varying from Bethe ansatz~\cite{shen_fan_prl,shen_fan,Bethe,sahandprl} (when the interactions between photons are local in character) to effective field theories \cite{bienas,Efimov,magrebi} (particularly successful for Rydberg nonlinear media).
Most of these techniques have been exclusively applied to few excitations and/or moderate size systems and, generally, the many-excitation dynamics remains largely unexplored. Recently an important step toward this direction 
was made in the specific case of an atomic array chirally (unidirectionally) coupled to a photonic waveguide. There, it has been recently demonstrated~\cite{mahmo_calajo} that a linear combination of quantum 
many-body photon bound states leads to the emergence  of the well-known self induced transparency (SIT) soliton~\cite{McCall1,McCall2,Bullough}. However, besides this specific case, a general connection  between the  quantum many-body dynamics and the emergent semi-classical behavior in other  atom-light interfaces is still generally unknown.


In recent years, an alternative formalism to capture quantum atom-light interactions has gained attention. This formalism is based on the insight that the light-matter polaritons that form in near-resonant propagation are in fact almost entirely atomic in character in typical settings. Thus, the photonic degrees of freedom can be effectively integrated out, to yield a quantum spin model describing photon-mediated interactions between the atoms \cite{anapra,anaprx,ana_masson,rev_wqedarray}. In principle, if the dynamics of the spin model can be solved, the correlations of the outgoing quantum fields can be readily obtained using an input-output equation \cite{Caneva,MPSJames,Caneva, MPSJames,glauber,Blais_in}. Within this framework, particularly in the context of atomic arrays and in waveguide QED, the spin model has led to the prediction of many interesting effects, including strongly subradiant states~\cite{Albrecht,Loic,ritch,dimermolmer1,zhang,buchler,dimer_shermet,Pod_scat,Pod_3,Lesanosky,Yudson}, photon mediated localized states~\cite{Pod_photon_loc}, topological states~\cite{Pod_topo}, chaotic states~\cite{caos} and subradiant dimers~\cite{dimermolmer2,dimer_poddu}. Techniques to effectively map continuous, macroscopic nonlinear media such as Rydberg ensembles to the spin model have also been proposed and investigated \cite{james_ryd}. 

Here, we utilize the spin model formalism to investigate photon bound states and their manifestation in photon propagation dynamics across different systems. In particular, we examine an interacting Rydberg atom ensemble, and also an array of two-level atoms coupled to a waveguide, treating both paradigmatic cases of bidirectional and chiral~\cite{Stannigel2012,Pichler2015,ChiralRev}  (unidirectional) emission. Within the spin model, the procedure to distinguish bound states  is analogous to that of identifying bound magnon excitations in condensed matter spin  models~\cite{Bethe,wortis,Schneider,Mattis,bloch,petroBS,cooper,ramos} and 
closely related to previous studies of dimers within waveguide QED~\cite{dimermolmer2,dimer_poddu,rev_wqedarray}. Here we first describe how the dispersion relation of two-photon bound states can be identified and then we show how these bound states manifest themselves in the spatio-temporal correlations of the output field, given classical weak incident pulses.
While the analytical techniques applied to two-photon bound states do not readily scale to investigate the properties of higher photon number, the spin model nonetheless admits an efficient way to numerically investigate the effect of many-photon bound states in the correlation of the output field given large incident pulses. This large photon number limit is made tractable through the possibility to naturally encode the spin model in matrix product state (MPS) representations. In this many-photon limit, we generally identify signatures of a transition  from true quantum many-body behavior, as characterized by non-trivial spatio-temporal correlations, to a semi-classical soliton wave reminiscent of self-induced transparency in atomic media, irrespective of the underlying details of the model. 

This paper is structured as follows. In Sec.~\ref{Sec. model}  we introduce a generic model for an ensemble of emitters coupled to a quasi-one-dimensional~(1D) quantum optical field. From here, we formulate the spin model, and present general criteria to identify the photon bound states. In Sec.~\ref{Sec.TLAarray} we consider as a first case an array of  two-level atoms~(TLA) coupled to either a  chiral or bidirectional  waveguide. We analyze the single- and two-excitation spectrum, followed by multi-photon propagation. In particular, in the many-photon limit, we recover the semi-classical phenomenon of self-induced transparency~(SIT). In Sec.~\ref{Sec.Rydmedia}, we apply the spin model to an ensemble of Rydberg atoms. After analyzing the two-excitation bound state dispersion relation, we make use of an effective two-level atom description to recover, in the many-body regime, a solitonic behavior similar to SIT, i.e. \emph{Rydberg SIT}. We conclude and provide an outlook of possible future directions in Sec.~\ref{sec:conclusions}.

\section{Model}\label{Sec. model}

\begin{figure}
\centering
\includegraphics[width=0.5\textwidth]{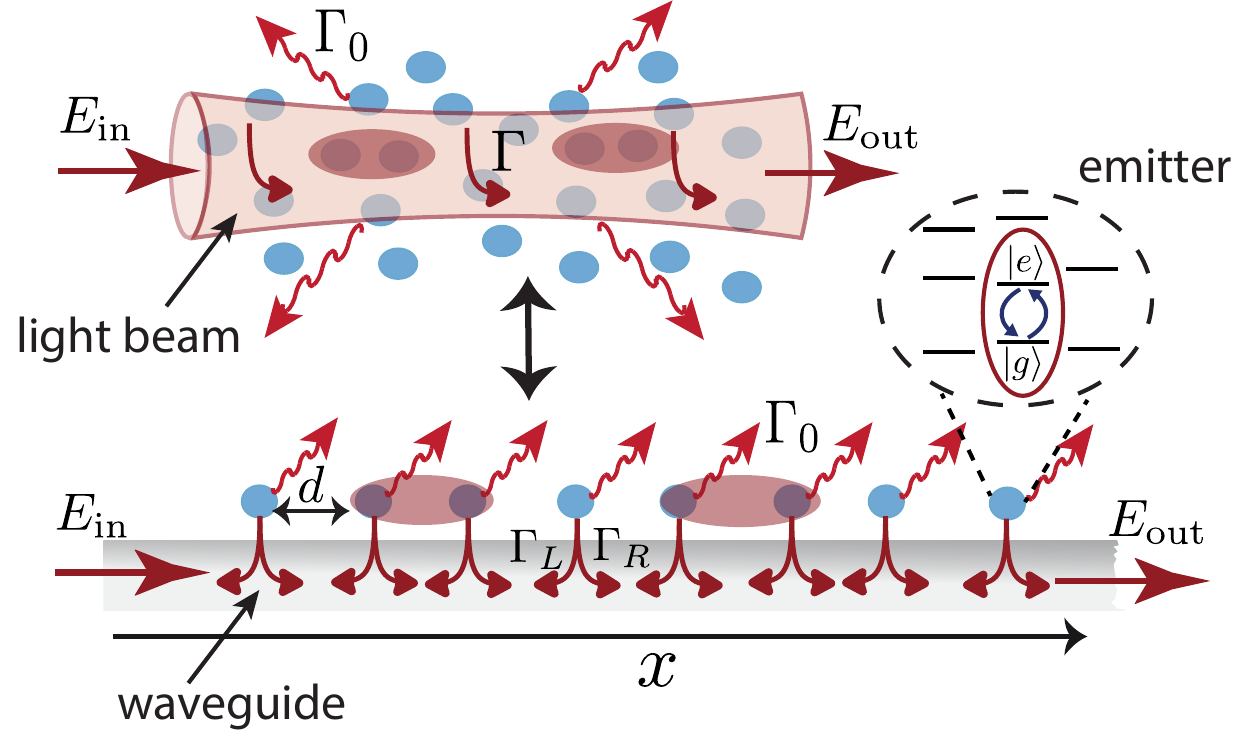}%
\caption{Generic quantum nonlinear media. A free-space ensemble of atoms with an unspecified level structure is coupled to photons in a focused propagating beam via a particular two-level transition. The system is characterized by a single-atom emission rate $\Gamma$ into the beam, and $\Gamma_0$ into other modes~(free space). Alternatively, we consider an atom array coupled to a one-dimensional optical waveguide. An individual atom emits into the left-propagating modes, right-propagating modes, and free space at rates $\Gamma_L, \Gamma_R$, and $\Gamma_0$, respectively. The free-space ensemble can be mapped to an effective waveguide model, as we discuss in the main text. In both scenarios, multi-photon bound states (red shaded area) can exist due to the nonlinear interactions induced by the atomic medium.}
\label{fig:setup}
\end{figure}

We consider a setup as illustrated in Fig.~\ref{fig:setup} where an ensemble of  $N$ atoms is coupled to a quasi-1D propagating field. We assume that the field couples to a transition of frequency $\omega_{eg}$ involving ground and excited states~$|g\rangle$ and $|e\rangle$, respectively, while the atoms can have other levels~(\textit{e.g.}, driven by classical control fields) that need not be specified now. The spontaneous emission rate of a single, excited atom into the 1D modes occurs at a rate $\Gamma$, while it can also emit at a rate $\Gamma_0$ into other modes besides the 1D continuum. This generic setup describes well a number of systems consisting of atoms coupled to a physical waveguide, where one might have $\Gamma_0\gg \Gamma$~(such as for atoms coupled to optical nanofibers)~\cite{Prasadnat,ReitzPRL2013,Mitsch2014,Pablo2017}, $\Gamma_0\sim \Gamma$~(for atoms coupled to photonic crystal waveguides)~\cite{Arcari2014,Sollner2015,Goban2014,Hood2016}, or $\Gamma_0\ll \Gamma$~(superconducting qubits coupled to unstructured or photonic crystal transmission lines)~\cite{Sundaresan,painter1,painter2,ustinov,ustinov_topo,marcoBS}. As pointed out in Ref.~\cite{MPSJames}, free space ensembles interacting with quasi-1D optical beams can also be mapped to this waveguide model, by taking $\Gamma_0\gg\Gamma$~(to account for the weak free-space light coupling) and finding suitable maps between the microscopic parameters of this model and the macroscopic parameters of the physical system such as optical depth~(discussed in detail in Sec.~\ref{Sec.Rydmedia}). For our situations of interest, we can consider the waveguide-coupled atoms to be in an array of lattice constant $d$, as shown in Fig.~\ref{fig:setup}. For bidirectional waveguides, this avoids Anderson or many-body localization of light associated with disorder and multiple scattering~\cite{Anderson,ReviewManybodyloc,NikosManybodyloc}. For chiral waveguides (and in free space, where the weak atom-light coupling is essentially equivalent to chiral coupling), the physics is in fact independent of the specific atomic positions, as we will describe below.

\subsection{Full Hamiltonian}
 Within the 1D model the full waveguide QED Hamiltonian of the system is given by the following contributions 
\begin{equation}\label{H_tot}
\hat H=\hat H_{ph}+ \hat H_a+\hat H_{in}.
\end{equation}
The fist term $\hat H_{ph}$ is the photonic Hamiltonian describing right- and left-propagating modes traveling in the channel with group velocity $c$.
It explicitly reads  ($\hbar=1$) 
\begin{equation}\label{H_ph}
\hat H_{ph}=-i\int dx\left[\hat E_R^{\dagger}(x)\frac{\partial}{\partial x}\hat E_R(x)-\hat E_L^{\dagger}(x)\frac{\partial}{\partial x}\hat E_L(x)\right]
\end{equation}
where $\hat E_{R(L)}(x)$ is the bosonic field operator annihilating a right-going (left-going) photon at position $x$ and fulfilling the commutation rule  $[\hat E_{R(L)}(x),\hat E^{\dagger}_{R(L)}(x')]=c\delta(x-x')$. 
$\hat H_a$ is the atomic Hamiltonian which at this stage is not fully specified. In particular, we will assume that the atom has ground and excited states $|g,e\rangle$ whose transition of frequency $\omega_{eg}$ couples to the quantum propagating field. Beyond that, the atom could contain additional levels, dissipation from other photonic channels (encoded in the rate $\Gamma_0$),  auxiliary control fields acting on different transitions, or other terms independent from the interaction of atoms with the quantum propagating field. The atom-light interaction allows for the creation or annihilation of excitations on the $n$-th atom through the spin operators $\hat\sigma^n_{ge}=|g_n\rangle\langle e_n|$ and $\hat\sigma^n_{eg}=|e_n\rangle\langle g_n|$. The corresponding interaction Hamiltonian reads
\begin{equation}\label{H_in}
\hat H_{in}=\sum_n\left[\hat\sigma_{ge}^n\left(\sqrt{\Gamma_R}\hat E_R^{\dagger}(x_n)+\sqrt{\Gamma_L}\hat E_L^{\dagger}(x_n)\right)+ \rm H.c.\right]
\end{equation}
where $\Gamma_R$ and $\Gamma_L$, with $\Gamma= \Gamma_R+\Gamma_L$, are the single-atom decay rates associated
respectively to the emission of right- and left-propagating photons. 
Here we explicitly distinguish the emission into the two directions to cover both the paradigmatic cases of bi-directional emission and chiral (uni-directional) emission that will be discussed in the rest of the paper.\\

\subsection{Spin model}\label{Secspin}

Standard treatments to tackle Hamiltonian~\eqref{H_tot} in the multi-excitation sector~(multiple photons and/or excited atoms) usually follow two different strategies depending on the regime considered. First, in the limit of strong dissipation, $\Gamma_0\gg\Gamma$, individual atoms have negligible interaction with light and interesting phenomena instead arise by collective coupling. Then, the ensemble of atoms can be treated as a continuous bosonic field. Interesting nonlinearities, such as arising from Rydberg interactions, can be added and be treated by effective field theories~(so far, limited to a few excitations)~\cite{bienas,Efimov,magrebi}, for example. Conversely, in the waveguide QED regime with negligible loss, $\Gamma_0\sim 0$, the eigenstates can be exactly computed using scattering theory formalism (e.g. S-matrix, Bethe ansatz, etc.)~\cite{shen_fan_prl,shen_fan,Bethe,sahandprl} for a few excitations, either for small atom number in general or for large atom number and chiral waveguides.

To go beyond specific limitations on excitation number or system details, we consider an alternative approach, where the photons are integrated out to arrive at an effective spin model for the atoms. This approximation takes advantage of the fact that in most physical systems of interest, the atom-photon dynamics occur on a time scale longer than the photon propagation time through the system, i.e. $\Gamma\ll c/Nd$, so that the light-mediated interactions can be considered instantaneous. Equivalently, the highly dispersive nature of the atoms due to the small linewidth causes the light-matter polaritons to be almost entirely atomic in character. The full system evolution is then given by the master equation (ME) for the reduced atomic density operator~\cite{Caneva, MPSJames,glauber} 
\begin{equation}\label{eq:MasterEq}
\dot{\hat \rho}= -i\left[ (\hat H_{\rm eff}+\hat H_{\rm drive})\hat \rho-\hat\rho (\hat H_{\rm eff}+\hat H_{\rm drive})^{\dagger}\right]+\mathcal{J}[\hat\rho].
\end{equation}
Here the non-Hermitian collective evolution
of the system is given by the effective Hamiltonian
\begin{equation}\label{Heff}
\hat H_{\rm eff}=\hat H_a-i\sum_{nm}A_{nm}\hat\sigma^n_{eg}\hat\sigma^m_{ge},
\end{equation}
where the matrix 
\begin{equation}
A_{nm}=\left[\Gamma_L e^{ik_0|x_n-x_m|}\theta_{nm}+\Gamma_R e^{ik_0|x_n-x_m|}\theta_{mn}\right]
\end{equation}
encodes the photon mediated atom-atom interactions and $\theta_{nm}=\theta(x_n-x_m)$, with $\theta_{nn}=1/2$, is the Heaviside function.  The Hamiltonian $\hat H_{\rm eff}$ is invariant 
under discrete translations of the product of the resonant wavevector $k_0=\omega_{eg}/c$ and the lattice constant $d$, $k_0 d \rightarrow k_0 d + 2\pi$, so a full description is obtained by considering $k_0d\in[-\pi,\pi]$.
The population recycling contribution to the evolution in Eq.\eqref{eq:MasterEq} is given by
 \begin{equation}
\mathcal{J}[\hat\rho]=\Gamma_0\sum_n\hat\sigma^{n}_{ge}\hat\rho \hat\sigma_{eg}^n+\sum_{nm}\left[\left(A_{nm}+A^*_{nm}\right)\hat\sigma^{n}_{ge}\hat\rho\hat\sigma_{eg}^m+\rm H.c.\right].
\end{equation}
Finally in Eq.\eqref{eq:MasterEq} we also explicitly add a driving term
\begin{equation}
\hat H_{\rm drive}=\sum_n\sqrt{\Gamma_R}\left[E_{in}(t,x_n)\hat\sigma_{eg}^n+ \rm H.c.\right],
\end{equation}
which couples the emitters to a right-propagating coherent state input field $E_{in}(t,x)=\mathcal{E}_{in}(t)e^{ik_{0} x-i\omega_{\rm in}t}$ with $\omega_{\rm in}$ being the central frequency of the driving field.\\

While such master equations describing photon-mediated dipole-dipole interactions have long existed, in recent years it has been realized that one can also use their solutions to re-construct the quantum field that has been previously integrated out, thus constituting a complete model of atom-light interactions. This field takes the form of an input-output relation~\cite{Caneva, MPSJames,glauber,Blais_in} 
 \begin{equation}\label{in_outR}
 \hat E_R(t)=\mathcal{E}_{\rm in}(t)+i\sum_n\sqrt{\Gamma_R}e^{-ik_0x_n}\hat\sigma_{ge}^n(t),
\end{equation}  
 \begin{equation}\label{in_outL}
 \hat E_L(t)=i\sum_n\sqrt{\Gamma_L}e^{ik_0x_n}\hat\sigma_{ge}^n(t).
\end{equation}  

In this work, we will on one hand consider the microscopic properties of the photon bound states themselves, which can be derived from the eigenstates of~\eqref{Heff} alone~(practically in the few-excitation limit). Separately, we will solve the full master equation dynamics of Eq.~(\ref{eq:MasterEq}) in the presence of the input field, to see how the bound states manifest themselves in the outgoing field properties of Eq.~(\ref{in_outR}). This dynamics can be computed numerically even in the many-excitation limit, with an MPS-based quantum trajectories algorithm discussed in Appendix~\ref{AppMPS}, and, as we are going to discuss in the following, is able to capture the progressive transition from a correlated output field to  semi-classical solitons. 
While we will specifically focus on coherent state (photon number uncertain) input fields here, these calculations can also be used to study Fock state inputs. 
 In particular, by simulating the dynamics with a quantum jump algorithm, one can post-select on the total number of jumps that occurred in the output fields to obtain a fixed photon number~\cite{MPSJames}.

 \subsection{Photon bound states}
Photon bound states are states with spatially correlated positions, which propagate through the bulk of a nonlinear medium experiencing low distortion~\cite{mahmo_calajo}. 
Despite the name, these bound states are actually almost entirely atomic~(spin-like) in nature when they propagate through the medium~\cite{chang_mirror,chang_tao}. The problem of identifying these states then becomes similar to that of magnon bound excitations in spin chains~\cite{Bethe,wortis}. For example, within the two-excitation subspace we can follow standard procedures to diagonalize the Hamiltonian~\eqref{Heff} in the relative and center-of-mass coordinate frame, which is convenient because the latter is characterized simply by a plane wave of wavevector $K$. On the other hand, the problem in the relative coordinate reduces to finding single-particle bound states in an effective (and possibly non-trivial) impurity model, whose energies depend parametrically on $K$. Due to the nonlinear interaction, this bound state will have a different energy $E_{\rm BS}$ than the sum of individual photon energies, which allows for their spectral identification:
 \begin{equation}\label{BS condition}
E_{\rm BS}(K)\ne J(q)+J(K-q).
\end{equation}
Here, $J(k)$ denotes the single-excitation dispersion relation and $q$ the relative momentum.

A crucial difference compared to canonical condensed matter spin models lies in the dissipative, long-range nature of the interactions encoded in the non-Hermitian Hamiltonian~\eqref{Heff}. A number of studies~\cite{dimermolmer1,dimermolmer2,dimer_shermet,dimer_poddu} have already pointed out that in a finite system, this Hamiltonian can give rise to two-excitation dimers as eigenstates that are generally lossy, physically due to the non-zero support of the wave functions with the system boundaries and their subsequent radiation into the empty waveguide. While these studies largely focused on how long-lived or subradiant these states could be, here, we more generally will study the dispersion relation of this continuum of bound states, and show that this dispersion relation and the spatial properties of the bound states indeed manifest themselves in the quantum nonlinear optics problem, in terms of correlations of the output field given a multi-photon input field.

 \subsection{Self-induced transparency}\label{Sec.SIT}

In~\cite{mahmo_calajo} it has been rigorously proven that in the many-body limit with finite size systems a linear combination of multi-photon bound states gives rise to the formation of self-induced transparency~(SIT) solitons, when the system consists of two-level atoms coupled to a chiral waveguide. As one of our main goals is to investigate whether SIT emerges in other quantum nonlinear systems from a full quantum picture, we first briefly review the effect of SIT, in the previously introduced language of waveguide QED.

SIT is a semi-classical phenomenon, involving the emergence of a soliton when the atoms are treated as two-level systems and the field classically. 
In order to write down the SIT equations, we move to a continuum description, mapping the spin operators to a continuous density, i.e. $\hat \sigma^n\rightarrow \hat \sigma(x)/\nu$, where $\nu$ is the linear density of the medium  and the spin operators fulfill the commutation relation $[\hat \sigma_{ge}(x),\hat \sigma_{eg}(x')]=-\hat \sigma_{z}(x)\delta(x-x')$ with $\hat \sigma_{z}(x)$ being the Z Pauli matrix. With this mapping, and assuming that the atoms resonantly interact with a right propagating field (we omit the $R$ label),
the total Hamiltonian~\eqref{H_tot} becomes
\begin{equation}\label{H_tot_cont}
\hat H=\!-i\!\!\int\!\! dx\hat E^{\dagger}(x)\frac{\partial}{\partial x}\hat E(x)+\sqrt{\Gamma}\!\!\int\!\! dx\left[\hat\sigma_{ge}(x)\hat E^{\dagger}(x)+ \!\rm H.c.\right].
\end{equation}

Using the Heisenberg equations with respect to this Hamiltonian and taking the expectation values $E(x,t)=\langle \hat E(x,t)\rangle$ and $\sigma(x,t)=\langle \hat \sigma(x,t)\rangle$, we get the mean field equations
\begin{equation}\label{eqMF}
\begin{split}
&\left[\frac{\partial}{\partial t}+c \frac{\partial}{\partial x}\right ]E(x,t)=-ic\sqrt{\Gamma}\sigma_-(x,t)\\
&\frac{\partial}{\partial t}\sigma_-(x,t)=i\sqrt{\Gamma}\sigma_z(x,t)E(x,t)\\
&\frac{\partial}{\partial t}\sigma_z(x,t)=4\sqrt{\Gamma }{\rm Im}[E(x,t)\sigma^*_-(x,t)],
\end{split}
\end{equation}
where  the external dissipation has been set to zero, $\Gamma_0=0$.
This set of equations~\eqref{eqMF} admits a solitonic solution for the photonic field (see Refs.~\cite{Bullough,mahmo_calajo} for a detailed derivation):
 \begin{equation}\label{eqSIT_solution}
  E(x,t)=\frac{ n_{\rm ph}\sqrt{\Gamma}}{2}\operatorname{sech}\left[\frac{ n_{\rm ph}\Gamma}{2}\left(\frac{x}{v_g}-t\right)\right],
 \end{equation}
with $v_g=(n^2_{\rm ph}\Gamma c)/(n^2_{\rm ph}\Gamma+4c\nu)$ being the group velocity inside the medium and $n_{\rm ph}$ being the average photon number in the pulse. Translated to a 1D   setting for a chain of $N$ atoms with linear density $\nu=1/d$, this leads to an overall pulse delay of~\cite{mahmo_calajo}
\begin{equation}\label{delaySIT}
\tau_{n_{\rm ph}}=4N/( n^2_{\rm ph}\Gamma).
\end{equation}
This solitonic solution has the property that its integrated Rabi frequency gives a 2$\pi$ pulse (area law)~\cite{McCall1,McCall2,Bullough,mahmo_calajo}
 \begin{equation}\label{eqSITarea}
2 \sqrt{\Gamma}\int dt E(x,t) = 2\pi,
 \end{equation}
letting each individual atom undergoing a full Rabi oscillation from the ground to excited state and back. This condition implies that photons are not taken away
from the original pulse, if the Rabi oscillation occurs on a time scale faster than the spontaneous emission rate $\Gamma_0$, allowing the soliton to propagate in a transparent fashion through the medium.



\section{Array of two-level atoms}\label{Sec.TLAarray}
As a first example of a quantum nonlinear medium, we consider an array of two-level atoms~(TLA) where the atomic Hamiltonian is simply given by $\hat H_a=(\omega_{eg}-i\Gamma_0/2)\sum_n\hat\sigma_{eg}^n\hat\sigma_{ge}^n$. In particular, we focus on the ideal waveguide QED scenario of low dissipation, $\Gamma_0\rightarrow 0$. In this regime there is a non-negligible probability that two propagating photons can interact simultaneously with the same atom  inducing a nonlinear optical response. Motivated by the previous discussion, we will first calculate the single-excitation dispersion relation, in order to identify energy gaps in which two-excitation bound states might exist, and  proceed to calculate their spectrum. We then consider the many-excitation limit observing numerically the emergence of a propagating soliton.  We will consider both cases of a chiral waveguide coupling, $\Gamma_L=0$ and $\Gamma_R=\Gamma$, where the results for multiple excitations are exactly known from the Bethe ansatz for photons~\cite{mahmo_calajo}, and a bidirectional waveguide, $\Gamma_L=\Gamma_R=\Gamma/2$.

\subsection{Single excitation sector}

\subsubsection{Chiral array}

We work in a rotating frame in order to discard the atomic frequency $\omega_{eg}$ and we assume  no additional dissipation~($\Gamma_0=0$). 
For chiral coupling, one can readily observe that the specific positions of the atoms do not affect the physics, provided that no two atoms are at the same position. In particular, in the master equation \eqref{eq:MasterEq} and in the input-output equations \eqref{in_outR} and  \eqref{in_outL}, one can make the transformation $\hat \sigma_{ge}^ne^{ik_0x_n}\rightarrow\hat \sigma_{ge}^n$~\cite{Pichler2015,ChiralRev} that eliminates all of the position-dependent phases $e^{ikx}$. For example, the effective Hamiltonian~\eqref{Heff} after this transformation reads:
\begin{equation}\label{Heffchi}
\hat H_{\rm eff}=-\frac{i}{2}\Gamma\sum_{n}\hat\sigma^n_{eg}\hat\sigma^m_{ge}-i\Gamma\sum_{n>m}\hat\sigma^n_{eg}\hat\sigma^m_{ge}.
\end{equation}
The second term~(involving the sum $n>m$) enforces that a given atom does not couple to atoms to the right, and thus encodes the unidirectional propagation of excitations along the array. 
For an infinite lattice of atoms ($N\rightarrow \infty$), the single-excitation sector is diagonalized by Bloch waves of wavevector $k$, $\hat H_{\rm eff}|\psi_k\rangle=J_k|\psi_k\rangle$ and the dispersion relation $J_k$ is given by 
\begin{equation}\label{dis_chi}
J_k=-\frac{\Gamma}{2}\rm cot\left[\frac{kd}{2}\right].
\end{equation}
The dispersion relation is plotted in Fig.~\ref{fig:setup_dis1ex}(a) and is characterized by a positive group velocity, $v^{(1)}_g(k)=\Gamma d/(4\sin^2{\left(\frac{kd}{2}\right)})$, for every value of $k$.
Interestingly, Eq.~(\ref{dis_chi}) cannot be derived by directly diagonalizing the Hamiltonian~\eqref{Heffchi} for a finite system, as $H_{\rm eff}$ is a triangular matrix with trivial eigenvalues. Instead, one should realize that the dispersion relation in principle should be real, and thus diagonalize the Hermitian part of the effective Hamiltonian:
\begin{equation}\label{Heff_herm}
\hat H_{\rm h}=-\frac{i\Gamma}{2}\left(\sum_{n>m}\hat\sigma^n_{eg}\hat\sigma^m_{ge}-\sum_{n>m}\hat\sigma^m_{eg}\hat\sigma^n_{ge}\right).
\end{equation}
This observation will be useful to numerically find the dispersion relation of multi-excitation bound states.

\subsubsection{Bidirectional ordered array}

\begin{figure}
\centering
\includegraphics[width=0.48\textwidth]{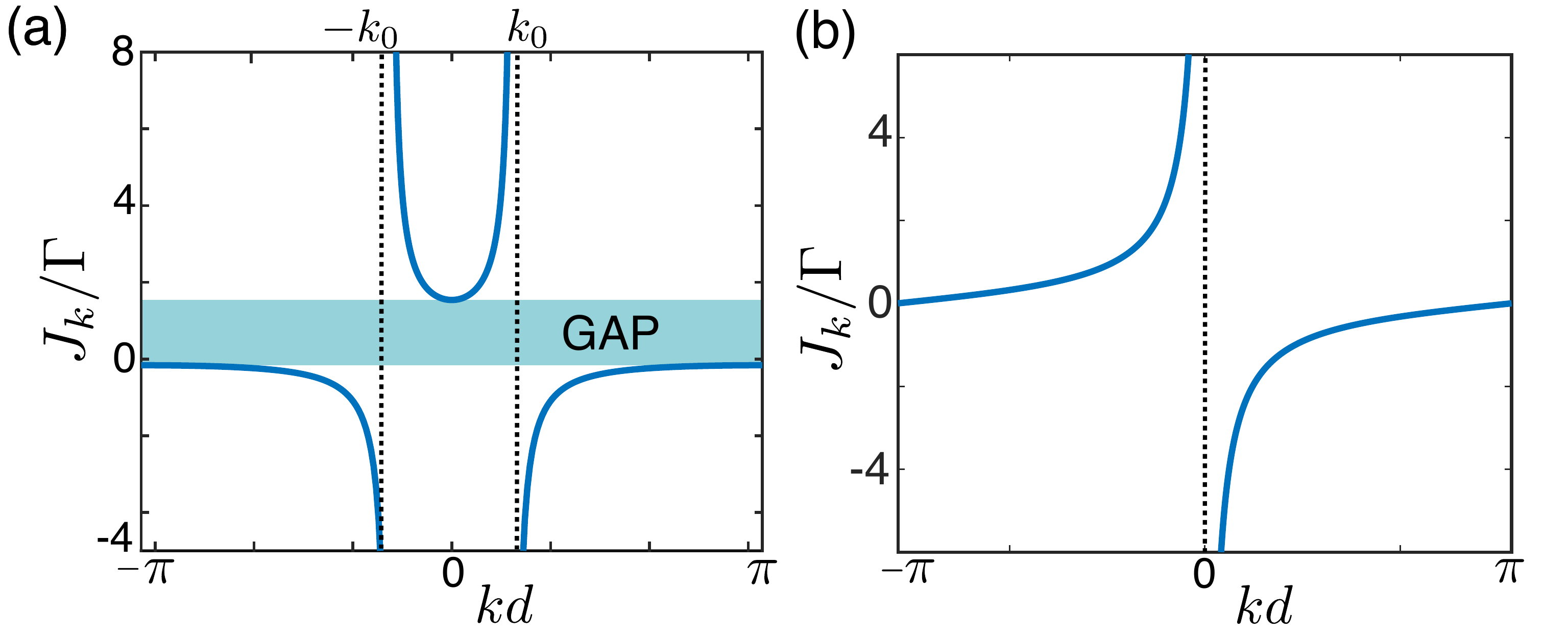}%
\caption{Single excitation dispersion relation $J_k$~(in units of the emission rate $\Gamma$) versus dimensionless wavevector $kd$ for the (a) chiral waveguide case, and  (b) the bidirectional case with $k_0d=0.2$. The vertical dashed lines indicate the values of the wavevector where the dispersion relation diverges.}
\label{fig:setup_dis1ex}
\end{figure}

For a bidirectional ordered array  Eq.~\eqref{Heff} reads
\begin{equation}\label{Heffbidi}
\hat H_{\rm eff}=-\frac{i}{2}\Gamma\sum_{mn}e^{ik_0|x_m-x_n|}\hat\sigma^n_{eg}\hat\sigma^m_{ge}.
\end{equation}
An infinite lattice is again diagonalized by Bloch waves and the  dispersion relation
 has previously been calculated to be~\cite{Albrecht}
\begin{equation}
J_k=\frac{\Gamma}{4}\left(\rm cot\left[\frac{(k_0+k)d}{2}\right]+\rm cot\left[\frac{(k_0-k)d}{2}\right]\right),\label{eq:Jk_bidirectional}
\end{equation}
as plotted in Fig.~\ref{fig:setup_dis1ex}(a). 
The dispersion relation~\eqref{eq:Jk_bidirectional} exhibits two distinct branches, along with a band gap near the atomic resonance frequency where no excitations are allowed. The dispersion relation reveals the polaritonic nature of the excitations, being close to the atomic resonance frequency~($J_k\sim 0$) for wavevectors significantly different than the resonant wavevector of light~($|k|\neq k_0$), while strongly hybridizing with light around $|k|\approx k_0$. Due to the Markov approximation, the slope of the polariton bands approaches the dashed vertical lines $|k|=k_0$ (see Fig.~\ref{fig:setup_dis1ex})(b) rather than a line of slope $c$. Note that the same behavior was also occurring in the chiral case with the dispersion diverging at $k=0$.

Notably, the dispersion relations of Eq.~\eqref{dis_chi} and Eq.~\eqref{eq:Jk_bidirectional} are purely real. This reflects the fact that in an infinite atom-waveguide system with no additional dissipation, the full system Hamiltonian \eqref{H_tot} (absent the input field) is Hermitian and thus the system forms lossless polaritons. Separately, we note that a large body of work has been devoted to the study of super- and sub-radiant eigenstates of finite atomic chains coupled to waveguides\cite{Albrecht,Loic,buchler}. In this case, the superradiant eigenstates are quasi-spin wave excitations with wavevectors close to the resonant wavevector of the waveguide ($k=k_0$ for bi-directional, and $k=0$ for chiral). As $N$ increases, the distribution of wavevectors that are superradiant becomes increasingly narrow, thus providing consistency with the lossless dispersion in the infinite system limit.



\subsection{Two excitation sector and bound states}\label{Secarray2ex}

Similar to the single-excitation sector, we can look for solutions of $\hat H_{\rm eff}|\psi^{(2)}\rangle=E|\psi^{(2)}\rangle$, where the two-excitation eigenstate generically takes the form $|\psi^{(2)}\rangle=\sum_{m<n}c^{(2)}_{mn}|e_m,e_n\rangle$. It is convenient to re-parametrize the coordinates in terms of center-of-mass~($x_{\rm cm}=(x_m+x_n)/2$) and relative coordinates~($x_r=|x_n-x_m|$). 
We will again consider the infinite system limit, where the dispersion relation should be purely real, and utilize Bloch's theorem to write the center-of-mass wave function in terms of a wavevector $K$,
\begin{equation}\label{2 ans}
 |\psi^{(2)}\rangle=\sum_{x_{\rm cm}}e^{iKx_{\rm cm}}f(x_r)|x_{\rm cm}-\frac{x_r}{2},x_{\rm cm}+\frac{x_r}{2}\rangle,
\end{equation}
where $f(x_r)$ is a generic function of the relative coordinate. This form allows one to reduce the two-excitation problem to a single-excitation one involving just the relative coordinate, and derive its spectrum as a function of $K$.

\subsubsection{Chiral array}\label{Sec_chiral2ex}

\begin{figure}
\centering
\includegraphics[width=0.48\textwidth]{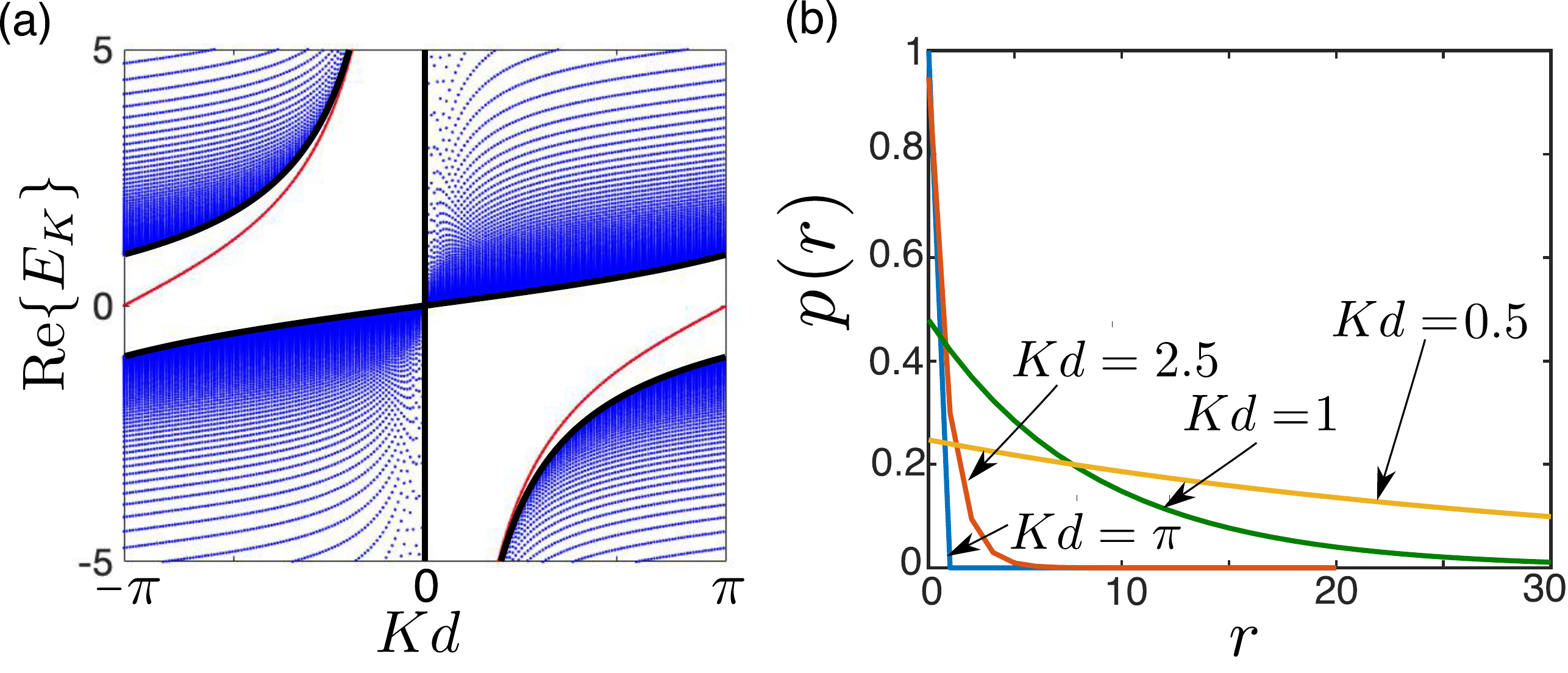}%
\caption{(a) Spectrum $E_K$ of two-excitation eigenstates, as obtained from Eq.~\eqref{HK_chiral}, as a function of the center of mass momentum $K$. The red curves indicate the bound state dispersion relation, while the blue dots indicate continuum states whose boundaries are given by the black curves. (b) Population distribution of the bound state as a function of the dimensionless relative coordinate $r=x_r/d$, for different values of total momentum $Kd$. These numerical calculations were performed for $N=150$.}
\label{fig:chiral_varie}
\end{figure}

Let us first consider the chiral waveguide case. This scenario is solvable by Bethe ansatz incorporating both photons and atoms~\cite{mahmo_calajo}, so we just briefly review how the same results emerge from the spin model. In the relative coordinate frame  Hamiltonian~\eqref{Heff_herm} can be rewritten as:
\begin{equation}\label{HK_chiral}
\hat H^K=-i\frac{\Gamma}{2}\sum_{r,r'>0}\sum_{\epsilon=\pm1}\left[e^{-i\frac{K}{2}|x_r+\epsilon x_{r'}|}-e^{i\frac{K}{2}|x_r+\epsilon x_{r'}|}\right]\hat\sigma^+_{r}\hat\sigma^-_{r'},
\end{equation}
which depends parametrically on the center of mass momentum $K$.
The full two-excitation spectrum can be easily obtained 
by numerically diagonalizing~\eqref{HK_chiral}. The real part of the eigenvalues, ${\rm Re}(E_K)$, as a function of the center of mass momentum $K$ is plotted in  Fig.~\ref{fig:chiral_varie}(a).  We implement this by truncating the single-particle problem of Eq.~\eqref{HK_chiral} to a large, but finite set of sites, with $1\leq r,r' \leq N-1$ and diagonalizing the resulting $(N-1)\times (N-1)$ matrix. The center of mass momenta are sampled at discrete points $K=2\pi m_K/N$, with $m_K=-N/2,...N/2-1$ for even $N$.

\begin{figure*}[!t]
\includegraphics[width=1.8\columnwidth]{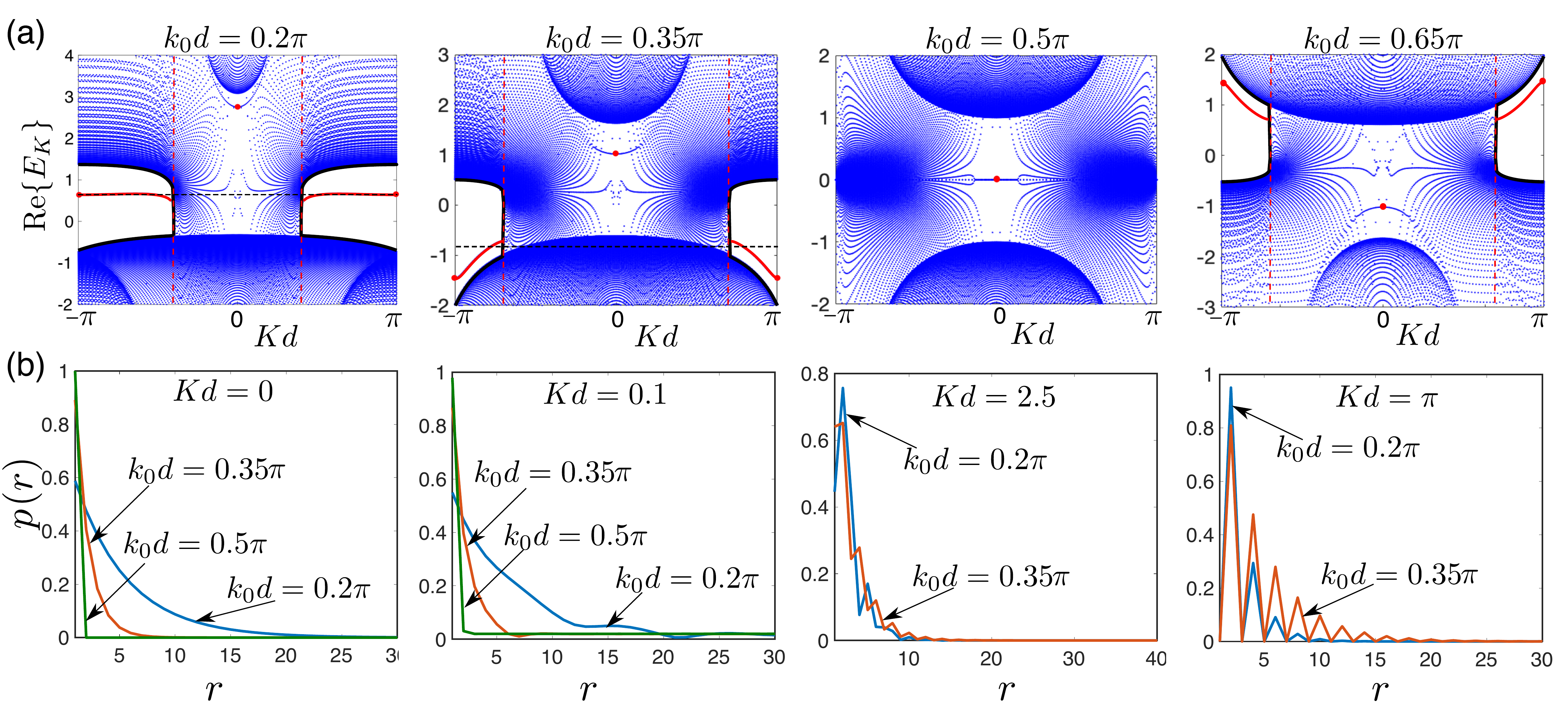}%
\caption{(a) Spectrum $E_K$ of two-excitation eigenstates of a bi-directional waveguide, as obtained from Eq.~\eqref{HK_bidi}, as a function of the center of mass momentum $K$ and for different values of the lattice constant $k_0 d$. The blue dots indicate continuum states, whose boundaries are denoted by the black curves. The solid red curves indicate the bound state dispersion relation, while the red dots indicate the special values $Kd=0$ and $Kd=\pm\pi$ where the bound-state properties can be computed analytically. The red dashed lines indicate the values of $K=\pm2k_0$ where the bound state dispersion branch emerges from the continuum. The horizontal dashed black lines in the first two panels correspond to twice the frequency of the input field used in Fig.~\ref{fig:pulse_bidi}. (b) Population distribution of the most localized state as a function of the dimensionless relative coordinate $r=x_r/d$, for different values of total momentum $Kd$ and lattice constant $k_0 d$. These numerical calculations were performed for $N=150$.}
\label{fig_bandK}
\end{figure*}

As anticipated in Sec.~\ref{Sec. model}, we observe in Fig.~\ref{fig:chiral_varie}(a) an energy continuum (blue eigenvalues) that corresponds to unbound states whose  energies can be obtained analytically by adding up the single-particle dispersion relation $\omega(K)=J(q)+J(K-q)$. 
 This continuum exhibits a gap where energy and momentum conservation are not simultaneously satisfied, whose boundaries are indicated by black curves.
Within this gap, we clearly observe a discrete dispersion branch~(red curve),  which  satisfies condition~\eqref{BS condition} and it can be identified as a two-excitation bound state.
This bound state exists for any value of $K$ and its dispersion relation can be 
exactly computed,  by using a Fourier transform ansatz, $|q\rangle=\sum_{r>0}e^{iqx_r}\hat\sigma^{r}_{eg}|0\rangle$,
\begin{equation}\label{dis_2ex}
E_{\rm BS}(K)=-2\Gamma\rm cot\left[\frac{Kd}{2}\right].
\end{equation}
The wave function itself is given by $p(r)\propto e^{-x_r \kappa}$ with $\kappa d=-i\log{\cos{\left(\frac{Kd}{2}\right)}}$, which is plotted in Fig.~\ref{fig_bandK}(b) for different values  of $Kd$.
The exact dispersion also allows to derive the bound state group velocity, 
\begin{equation}\label{vgchiral}
v^{(2)}_g(K)=\frac{\partial E_{\rm BS}(K)}{\partial K}=\frac{\Gamma d}{\sin^2{\left(\frac{Kd}{2}\right)}}=4v^{(1)}_g(k).
\end{equation}
In other words, the bound state travels faster than a single excitation. This result coincides with the Bethe ansatz calculation of Ref.~\cite{mahmo_calajo}, which is not surprising, since the polariton propagation speed is almost entirely dictated by the atomic dispersion, rather than the speed of light itself.

\subsubsection{Bidirectional ordered array}

Using the same methodology as in the chiral case, we rewrite Hamiltonian ~\eqref{Heffbidi}  in the relative coordinate frame:
\begin{equation}\label{HK_bidi}
\hat H^K=-i\frac{\Gamma}{2}\sum_{r,r'>0}\sum_{\epsilon,\epsilon'=\pm1}e^{i(k_0+\epsilon\frac{K}{2})|x_r+\epsilon'x_{r'}|}\hat\sigma^{r}_{eg}\hat\sigma^{r'}_{ge},
\end{equation}
and numerically compute the full spectrum, which is plotted in Fig.~\ref{fig_bandK}(a) for different values of the atomic distance $k_0d$. 

Similarly as in the chiral case, we observe a continuum of unbound states~(blue points) with energies given by $\omega(K)=J(k)+J(K-q)$ (black curve).
 On the other hand, a gap of forbidden energies occurs in the intervals $Kd\in[2k_0d,\pi]$ and $Kd\in[-\pi,-2k_0d]$ (red dashed lines), and at the singular point $Kd=0$. Note that the eigenvalues surrounding $Kd=0$ are characterized by a vanishing density of states but do not exhibit a gap. 
Within the gaps, we find some discrete two-excitation bound state energies that satisfy condition~\eqref{BS condition} and, besides the singular point at $Kd=0$, 
present a well-defined dispersion branches (red lines).
For the special values $Kd=0$ and $Kd=\pm\pi$ the two-excitation bound states have a simple analytical form~\cite{dimermolmer2},  which can be obtained  by using a Fourier transform ansatz in the relative coordinate.
These three dimers have energies $\omega_0=2\Gamma \rm cot(k_0d)$ and $\omega_{\pi}=\omega_{-\pi}=2\Gamma \rm cot(2k_0d)$, and
are characterized  by a complex relative  momentum, $q_0 d=\log{\cos{(k_0d)}}$ and $q_{\pm\pi}d=-\log{\cos{(2k_0d)}}$. These complex momenta  lead to an exponential localization of the wave function along the relative coordinate, which explicitly reads $p_0(r)\propto e^{-2x_r q_0}$ and $p_{\pm\pi}(r)\propto \cos^2( \frac{Kx_r}{2})e^{-x_r q_{\pm\pi}}$. Note that while for the dimer at $Kd=0$ the two excitations have a relative distance of $d$ given by the TLA saturation, for $Kd=\pm\pi$ they are separated by $2d$ due to the oscillating term in the population. An analytical expression for the bound states with generic center of mass momentum is reported in Ref.~\cite{Bakkensen}.
The population distribution of these states is shown in Fig.~\ref{fig_bandK}(b) as a function of the relative coordinate, for different representative values of the total momentum $Kd$ and of the atomic separation $k_0d$. A full localization is indeed observed when the parameter choices coincide with a gap. However, the neighborhood around $Kd=0$ does not exhibit a true bound state solution, as evidenced by the large-$r$ tail in the population of the most localized state plotted in Fig.~\ref{fig_bandK}(b) for $Kd=0.1$. These quasi-localized states in the continuum arise from scattering resonances and are extensively discussed in Ref.~\cite{Bakkensen}.

Such a bound state dispersion relation, as far as we know, has not been previously derived when the photonic degree of freedom is explicitly kept. The reduced complexity is one of the strengths of the spin model approach, which we will soon see also extends to numerically exploring the many-body limit.

\subsection{Many-body pulse propagation}\label{Sec_TLA_many}


While analytically deriving the properties of few-excitation bound states (beyond two) seems generally challenging within the spin model, this framework still offers a route towards numerical investigations, even in the many-body limit. 
In this section we utilize such numerics to explore the dynamics of these many-body bound states.  This problem has been recently studied in Ref.~\cite{mahmo_calajo} for an array of atoms coupled to a chiral waveguide, which established the connection between quantum photon bound states
 and the emergence of the SIT soliton discussed in Sec.\ref{Sec.SIT}. In the following, we first briefly summarize the general methodology and the results obtained for the chiral case and then we explore the nature of this transition in a different setting: a bidirectional array.

\subsubsection{Methodology}\label{Secmet}

In order to investigate the change in behavior from weak to strong pulses, we utilize a previously developed matrix product state (MPS) algorithm for this problem (see App.~\ref{AppMPS} and Ref.~\cite{MPSJames,james_ryd,mahmo_calajo} for further details). In this representation, the maximum bond dimension needed to obtain convergence in the simulations, $D_{\rm max}$, is directly connected to the entanglement entropy of the system and it acts as an important figure of merit to understand if the dynamics is highly correlated, $D_{\rm max}\gg1$, or if a mean field approximation for the atoms suffices,  $D_{\rm max}\sim 1$.

With this formalism we solve the full emitter dynamics, as governed by the master equation~\eqref{eq:MasterEq}. 
Considering that the finite spatial extent of the bound states generally allows for their excitation given a coherent state input pulse,
we specifically start with the atoms in the ground state $|g\rangle^{\otimes N}$ at $t=0$, and we consider two different  pulse shapes. The first consists of a Gaussian pulse, peaked at time $t_0$, with amplitude $\mathcal{E}_{\rm in}(t)=\sqrt{n_{\rm ph}}e^{(t- t_0)^2/(2 \sigma^2)}/(\sqrt{\sigma}\pi^{1/4})$, where $n_{\rm ph}$ is the average number of photons in the pulse and $\sigma$ the pulse width.  
The second, useful  to capture the solitonic transition, has an SIT-like solitonic shape~\cite{McCall1,McCall2,Bullough,mahmo_calajo},  $\mathcal{E}_{\rm in}(t)=(\tilde n_{\rm ph}\sqrt{\Gamma_R}/2)\operatorname{sech}(\tilde n_{\rm ph}\Gamma_R (t-t_0)/2)$,  that matches the 
SIT solution given in Eq.~\eqref{eqSIT_solution}. Here  $\tilde n_{\rm ph}=\sqrt{n_{\rm ph}^2+(\omega_{\rm in}/\Gamma)^2}$ is the generalized Rabi amplitude and we remind that for the chiral case $\Gamma_R=\Gamma$.

Once sent an incoming pulse trough the system, the transmitted intensity $I_R=\langle \hat E_R^{\dagger}(t) \hat E_R(t)\rangle$ and the equal-time higher-order correlation functions $G^{(m)}(t)=\langle \left[ \hat{E_R}^\dagger(t)\right]^m\left[\hat{E_R}(t)\right]^m \rangle$~($m\geq 2$) are
 computed by using the expression for the output field operator given in Eq.~\eqref{in_outR} in terms of the input field amplitude and emitter correlations.

\subsubsection{Many-body bound states-SIT transition in a chiral array}

\begin{figure}
\includegraphics[width=0.48\textwidth]{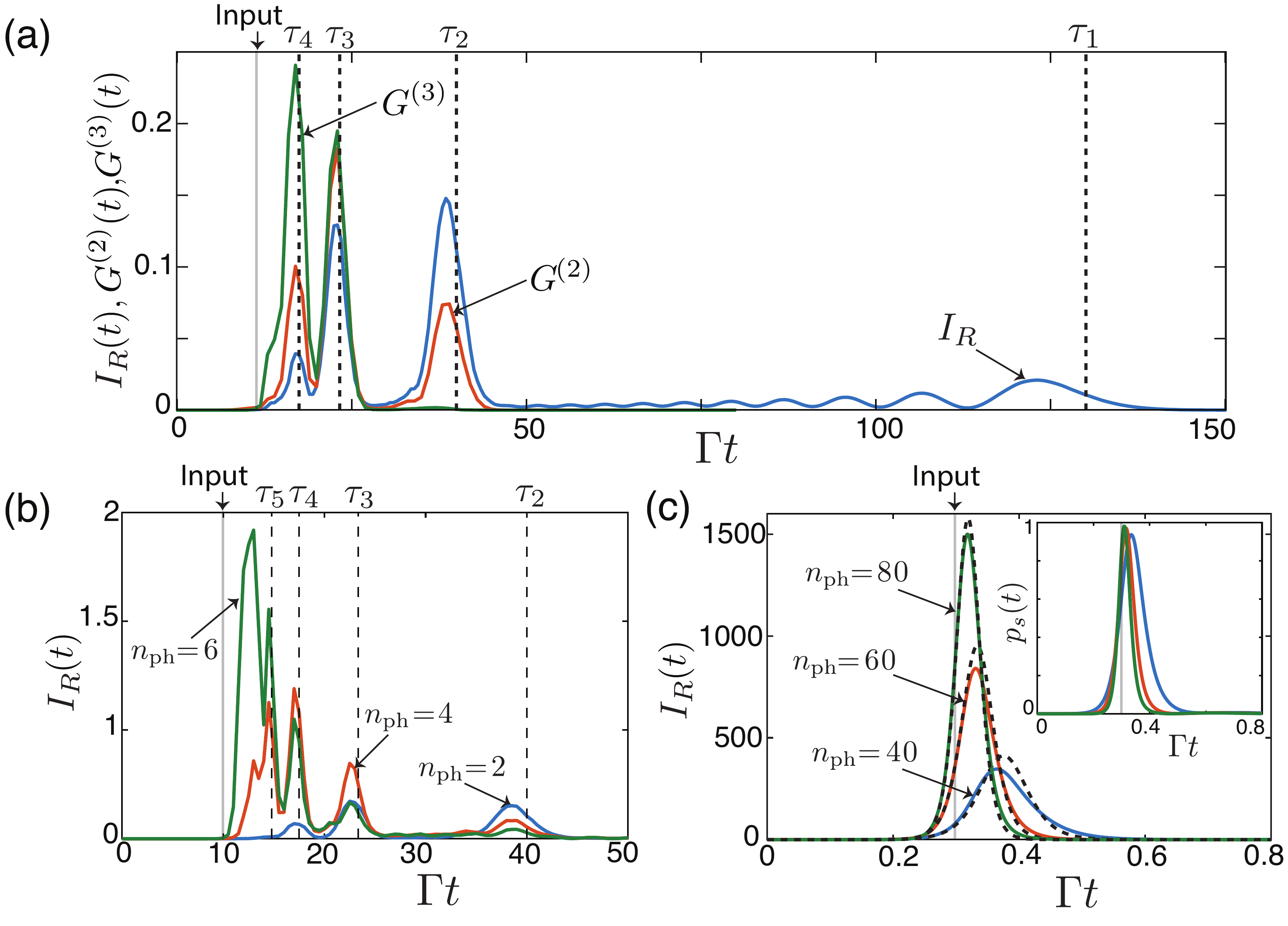}%
\caption{Few to many-body pulse propagation in a chiral  array of $N=30$ atoms. In (a) we plot the time-dependent output intensity $I_R(t)$ and correlation functions $G^{(n)}(t)$, given a coherent Gaussian input pulse with width $\sigma\Gamma=3$ and average photon number $n_{ph}=2.0$. The input frequencies are set to resonance with the atomic transition, $\omega_{\rm in}=0$.  In (b)-(c) we plot the output intensity $I_R(t)$ for a solitonic input pulse (see main text) with different strengths $ n_{\rm ph}$ as indicated in the panels.  In (c) the black dashed line corresponds to  the mean field SIT soliton solution. In all plots the peak of the input pulse is indicated by the grey vertical line and is set to $\Gamma t_0=10$ for (a)-(b) and  $\Gamma t_0=0.3$ for (c). The vertical dashed lines  indicate  the expected delay of the many-body photon bound states. In the inste of (c) we also plot the atomic population of an atom in the middle of the array  as a function of time for the same average photon number as in (c). The simulation has been performed with an MPS based quantum trajectories algorithm involving $N_t=5000$ trajectories and maximum bond dimension $D_{\rm max}=40$.}
\label{fig:pulse_chiral}
\end{figure}

For a chiral array, it has been proven that a coherent pulse resonant with the atomic transition can efficiently excite $n_{\rm ph}$-excitation bound states that experience a photon number-dependent time delay, $\tau_{n_{\rm ph}} = 4N/(n_{\rm ph}^2\Gamma)$. Note that for $n_{\rm ph}=2$, this result coincides with the bound state group velocity derived in Eq.~\eqref{vgchiral}. In sufficiently long systems, the combination of photon number uncertainty of the coherent state and the number-dependent delay results in a train of correlated photons ordered by photon number at the output~\cite{mahmo_calajo}, as shown in Fig.~\ref{fig:pulse_chiral}(a). Here we used an input Gaussian pulse, on resonance with the atomic transition and with $n_{\rm ph}=2$  and $\sigma\Gamma=3$, and we plotted   the output intensity and the photon correlation functions as a function of time, indicating with the vertical dashed line the expected delay.  Increasing the average photon number of the input pulse, large photon bound states with small delay are excited and the number components become progressively less separated. This is shown in Fig.~\ref{fig:pulse_chiral}(b)-(c) where we used the solitonic input pulse and we compared the ouptut intensities for different average photon numbers $n_{\rm ph}$. Note that here  the incoming pulse is on resonance with the atomic transition thus $\tilde n_{\rm ph}=n_{\rm ph}$. For sufficiently large photon number, the individual bound states cannot be resolved anymore and a single soliton-like peak emerges (see Fig.~\ref{fig:pulse_chiral}(c)). This behavior is also associated to a full $2\pi$ rotation of the atoms, as shown in the inset of Fig.~\ref{fig:pulse_chiral}(c), where we plot the  time-dependent excited-state population $p_s(t)$ of a single atom in the middle of the array.
In this case, it can be explicitly shown that a coherent state distribution of bound states of different photon number in fact coincides with the classical SIT soliton~\cite{McCall1,McCall2,Bullough,mahmo_calajo}, given in Eq. \eqref{eqSIT_solution} and plotted in Fig.~\ref{fig:pulse_chiral}(c) with the black dashed line, thus establishing a crossover from quantum to classical nonlinear optics. Indeed, one can note that the predicted time delay coincides with the one given in Eq.~\eqref{delaySIT} obtained by the solution of the SIT mean-field equations. 
While here, we have simply summarized the results already derived for two-level atoms coupled to a chiral waveguide, in the following, we would like to explore the nature of the transition to large photon number for other systems, and look for the emergence of soliton behavior.

\subsubsection{Exciting many-body bound states in a bidirectional array}
 
\begin{figure}
\includegraphics[width=0.48\textwidth]{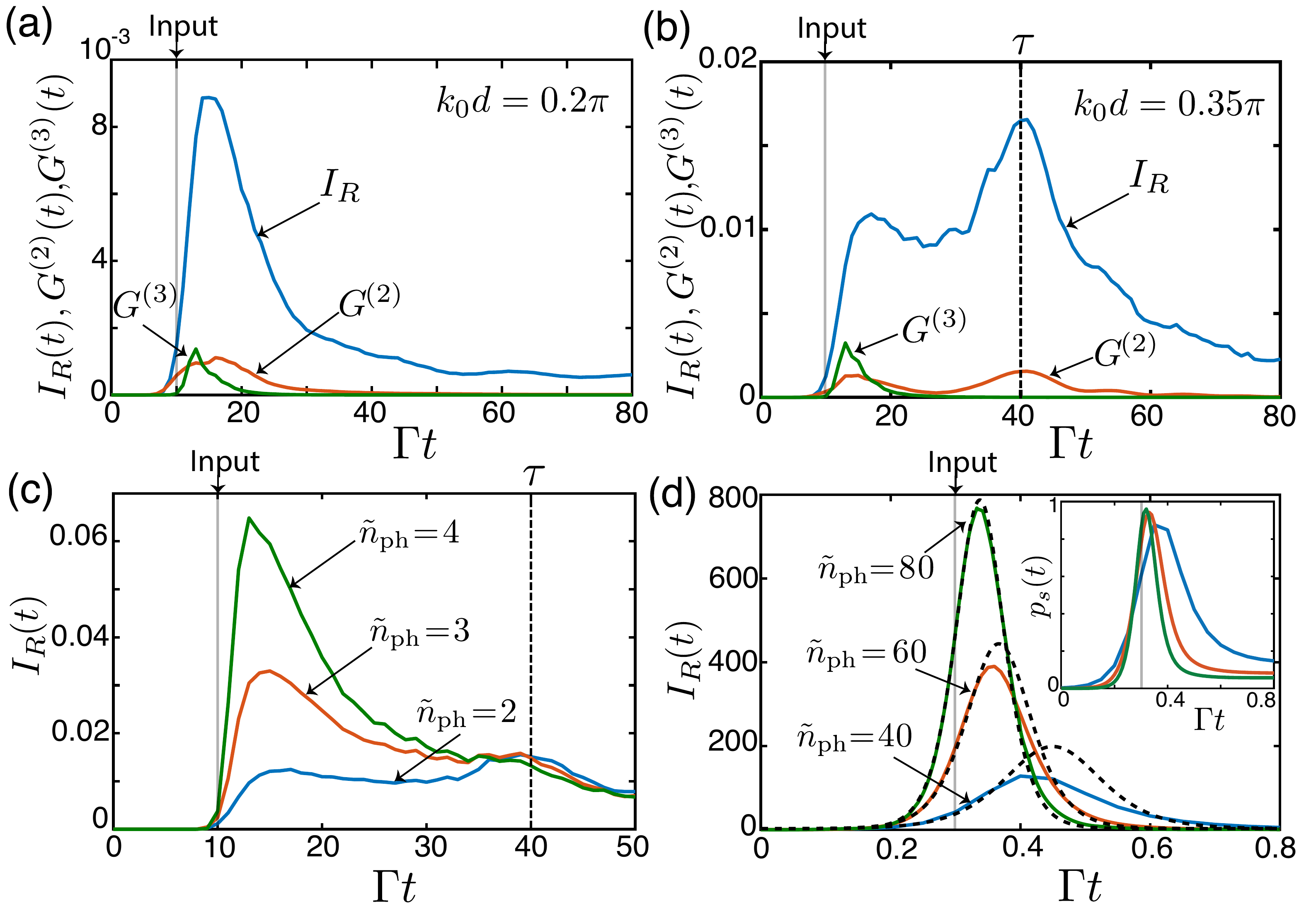}%
\caption{Few to many-body pulse propagation in a bi-directional array of $N=30$ atoms. In (a)-(b) we plot the time-dependent output intensity $I_R(t)$ and correlation functions $G^{(n)}(t)$, given a coherent Gaussian input pulse with width $\sigma\Gamma=3$ and average photon number $n_{ph}=2.0$, for two different values of the atomic distance $k_0d$. The input frequencies are respectively $\omega_{\rm in}=0.33$ in (a) and $\omega_{\rm in}=-0.4\Gamma$ in (b) and are indicated by the black horizontal dashed lines of Fig.~\ref{fig_bandK}(a). In (c)-(d) we plot the output intensity $I_R(t)$ for a solitonic input pulse (see main text) with different strengths $\tilde n_{ph}$ as indicated in the panels. Here we fix $k_0d=0.35\pi$ and $\omega_{\rm in}=-0.4\Gamma$. In (d) the black dashed line corresponds to  the mean field SIT soliton solution. In all plots the peak of the input pulse is indicated by the grey vertical line and is set to $\Gamma t_0=10$ for (a)-(c) and  $\Gamma t_0=0.3$ for (d). The vertical dashed lines in  (b) and (c) indicate  the expected delay of the two-photon bound states. In the inste of (d) we also plot the atomic population of an atom in the middle of the array  as a function of time for same average photon number as in (d).  The simulation has been performed with an MPS based quantum trajectories algorithm involving $N_t=5000$ trajectories and maximum bond dimension $D_{\rm max}=40$.}
\label{fig:pulse_bidi}
\end{figure}

We now explore the analogue of these effects in a bi-directional array. One can already see that a major difference compared to the chiral case is the emergence of a single-excitation band gap around the atomic resonance, as shown in Fig.~\ref{fig:setup_dis1ex}(a), which causes strong reflectance of weak resonant pulses. On the other hand, the experimental observation of SIT does not rely on chirality, and so one might expect that SIT will arise for strong enough input pulses in the bi-directional case. In the following, we investigate the how the behavior of the system changes going from weak to strong pulses.

We start by considering an input Gaussian pulse as the one used for the chiral case with $n_{\rm ph}=2$  and $\sigma\Gamma=3$.
Representative output intensities and correlation functions for such a setup are plotted in Fig.~\ref{fig:pulse_bidi}(a)-(b), for an array of $N=30$ atoms with lattice constants $k_0d=0.2\pi$ and $k_0d=0.35\pi$. Considering a large array is crucial because the propagating bound states experience a time delay at the output, which is proportional to the number of emitters and the inverse of their group velocity evaluated at the frequency of the input pulse $\omega_{\textrm{in}}$, i.e. $\tau=Nd/v_g(\omega_{\rm in})$. 
Note that the bound state energies occur at frequencies away from the atomic resonance, so we consider input fields with a slightly detuned central frequency, i.e. $\omega_{\textrm{in}}\neq \omega_{eg}$. For the lattice constant $k_0d=0.2\pi$, we set $\omega_{\textrm{in}}\approx 0.33\Gamma$, which is situated within the single-excitation band gap of Fig.~\ref{fig:setup_dis1ex}(a), which suppresses transmission of the single-photon component of the pulse. For the case of $k_0d=0.35\pi$, the chosen input frequency $\omega_{\textrm{in}}=-0.4\Gamma$ instead coincides with the flat, uppermost region of the lower dispersion branch, leading to a large delay of the transmitted single-excitation component.

In the case of inter-atomic distance $k_0d=0.2\pi$, the two-excitation bound state dispersion (Fig.~\ref{fig_bandK}(a)) is extremely flat and significantly deviates from a simple linear slope over the bandwidth of the incoming pulse. Thus, the excited bound states exhibit both a large delay and significant distortion at the output, which suppresses the appearance of a clear peak associated to the bound states in the output intensity of Fig.~\ref{fig:pulse_bidi}(a). One can also observe a faster exiting peak in the higher-order correlation $G^{(3)}(t)$, which could correspond to the excitation of higher photon number bound states. A similar peak immediately following the input is also observed in Fig.~\ref{fig:pulse_bidi}(b) for the lattice constant $k_0d=0.35\pi$. Importantly though, here we also resolve another intensity peak with a significant delay $\Gamma(\tau-t_0)\sim 30$, and an associated peak in $G^{(2)}(t)$. This peak can be clearly associated to the excitation of a two-photon bound state resonant with the input field, as indicated by the dashed black line in the second panel of Fig.~\ref{fig_bandK}(a). The numerically calculated group velocity $v_g(\omega_{\textrm{in}})$ at this frequency can be used to predict the expected delay (vertical dashed line in Fig.~\ref{fig:pulse_bidi}(b)), and coincides closely with the peaks in the intensity and $G^{(2)}$. 

In summary, as with the chiral case, it is possible to observe peaks in the output field and its correlations, which are associated with the excitation of few-photon bound states. However, unlike the chiral case, this observation requires more fine tuning in the bi-directional array, due to the finite bandwidth of the bound state band and its possibly strong group velocity dispersion.

\subsubsection{Towards the SIT limit}
We now investigate progressively higher photon number inputs, and the robust transition toward the mean field solution of SIT. Anticipating such a transition, we now drive the system with the SIT-like solitonic input field introduced in Sec.~\ref{Secmet}. This creates a large overlap between the field and the many-photon bound state, assuming the latter approaches the SIT solution. 

In Fig.~\ref{fig:pulse_bidi}(c)-(d) we plot the resulting output intensity for photon numbers ranging from small~($\tilde{n}_{\rm ph}=2$) to large. In  Fig.~\ref{fig:pulse_bidi}(c), for small $\tilde{n}_{\rm ph}$, we clearly observe the appearance of the two bound state peaks previously discussed. Not surprisingly, the second, more delayed peak associated with the two-photon bound state gets rapidly suppressed as the photon number is increased. For even larger photon numbers as plotted in Fig.~\ref{fig:pulse_bidi}(d), we see that the output intensity gradually approaches the mean field solitonic solution
\begin{equation}
\bar I_R(t)=\frac{\tilde n^2_{\rm ph}\Gamma_R}{4}\operatorname{sech}^2\left[\frac{\tilde n_{\rm ph}\Gamma_R(t-t_0-\tau_{\tilde n_{\rm ph}})}{2}\right],
\end{equation}
which is delayed respect to the input by an amount  $\tau_{\tilde n_{\rm ph}}=4N/(\tilde n^2_{\rm ph}\Gamma_R)$~\cite{mahmo_calajo}. The convergence toward this mean field solution also indicates that the medium becomes more transparent to the large photon number pulse, and that the impedance matching of the pulse to the atomic array becomes irrelevant despite the bidirectional coupling to the waveguide. 
As in the chiral case, the emergence of a transparent soliton is 
also signaled by  a full $2\pi$ rotation of the excited-state population $p_s(t)$, as shown in the inset of Fig.~\ref{fig:pulse_bidi}(d).
Finally, we comment that, similarly as  studied in Ref.~\cite{mahmo_calajo}, the transition toward classical behavior is evidenced by the decreasing maximum bond dimension required for convergence in the MPS simulations, from $D_{\rm max}\sim 40$, for $\tilde n_{\rm ph}\sim1$, to $D_{\rm max}\sim 1$, for $\tilde n_{\rm ph}\gg1$. 


\subsubsection{Experimental considerations}\label{experimentTLA}

The bound states studied here have the advantage of not relying on a chiral waveguide setup, which was the focus of Ref.~\cite{mahmo_calajo}.
Although schemes to realize chiral coupling with superconducting qubits coupled to microwave transmission lines has been proposed~\cite{Guimond,Gheeraert}, it is more routine to realize bi-directional coupling in circuit QED setups either to microwave transmission lines or coupled resonator arrays~\cite{Sundaresan,painter1,painter2,ustinov,ustinov_topo,marcoBS}. Such systems are ideal to investigate the physics predicted above, as the artificial qubits can experience extremely low loss into channels other than waveguide emission, $\Gamma_0/\Gamma<10^{-2}$, and can be located at precise positions along the waveguide.

\section{Photon bound states in Rydberg media}\label{Sec.Rydmedia}

We now consider a different quantum nonlinear medium consisting of a free space ensemble of Rydberg atoms. The level scheme of the atoms is pictorially shown in Fig.~\ref{Fig.Rtran}(a) where, besides the $|g\rangle$ and $|e\rangle$ states, there is an additional long-lived Rydberg level $|s\rangle$, which couples to $|e\rangle$ via a uniform, classical control field of Rabi frequency $\Omega$ and detuning $\delta$. The propagation of photons interacting with atoms on the $|g\rangle$-$|e\rangle$ transition can become highly nonlinear, due to strong van der Waals interactions between Rydberg excitations. In particular, the van der Waals interaction is responsible for shifting the resonant energy of a Rydberg level by an amount $V(x_r)=C_6/x_r^6$ given the presence of another Rydberg excitation nearby, with $x_r$ being the relative distance between the two atoms. This can significantly modify the propagation of two photons~(or more precisely, Rydberg polaritons) that are closer than a Rydberg blockade radius $r_b$, to be defined later. We first introduce the effective model used to describe the system, and then we numerically investigate the many-body photon dynamics. The results obtained by the full simulation are then interpreted in terms of an effective theory for a generalized version of SIT, which we show as being able to capture the main emergent features.


\subsection{Model and spectrum}\label{Sec.modelRyd}
The three-dimensional Rydberg ensemble can be approximately treated as one dimensional, provided that the blockade radius is larger than the beam waist of the photons exciting the $|g\rangle$-$|e\rangle$ transition, such that the photon dynamics occurs within a single transverse mode. As anticipated in Sec.~\ref{Sec. model}, this situation can then be mapped to a 1D array of atoms coupled to a chiral waveguide~\cite{bienas,MPSJames}, with a large additional and independent excited-state dissipation rate $\Gamma_0\gg \Gamma$ to capture the emission of photons into $4\pi$ modes other than the mode of interest. The chirality suppresses reflection from the array~(reproducing the negligible reflection of a free-space ensemble). We will later see how the artificial parameters of our microscopic model~(such as $N,\Gamma,\Gamma_0$) can be matched to physical, macroscopic quantities in an actual ensemble, and in particular, how a numerically tractable atom number $N$ and moderate $\Gamma_0$ for our system can be used to deduce the physics in an ensemble that exhibits much larger $N$ but simultaneously larger $\Gamma_0$.

Within the waveguide mapping, the corresponding spin model~(see  Sec.~\ref{Secspin}) in a rotating frame is given by
\begin{equation}\label{Haryd}
\begin{split}
\hat H_a=&-\frac{i\Gamma_0}{2}\sum_n\hat\sigma_{eg}^n\hat\sigma_{ge}^n -\delta\sum_n \hat\sigma^n_{sg}\hat\sigma^n_{gs}+\Omega\sum_n\left(\hat\sigma_{es}^n+\hat\sigma_{se}^n\right)\\
&+\sum_{n<m}V(|x_n-x_m|)\hat\sigma^n_{sg}\hat\sigma^n_{gs}\hat\sigma^m_{sg}\hat\sigma^m_{gs},
\end{split}
\end{equation}
where $\hat\sigma^n_{gs}=|g\rangle\langle s|$ ($\hat\sigma^n_{sg}=(\hat\sigma^n_{gs})^\dagger$) and $\hat\sigma^n_{es}=|e\rangle\langle s|$ ($\hat\sigma^n_{se}=(\hat\sigma^n_{es})^\dagger$)  are the operators associated with the $|g\rangle$-$|s \rangle$ and $|e\rangle$-$|s \rangle$ transitions of atom $n$, respectively, and $\delta=\omega_c-\omega_{se}$ is the detuning of the control driving field $\Omega$ with respect to the $e$-$s$ transition. Here, we assume that dissipation on the Rydberg level can be neglected owing to its long lifetime.
\begin{figure}
\includegraphics[width=0.48\textwidth]{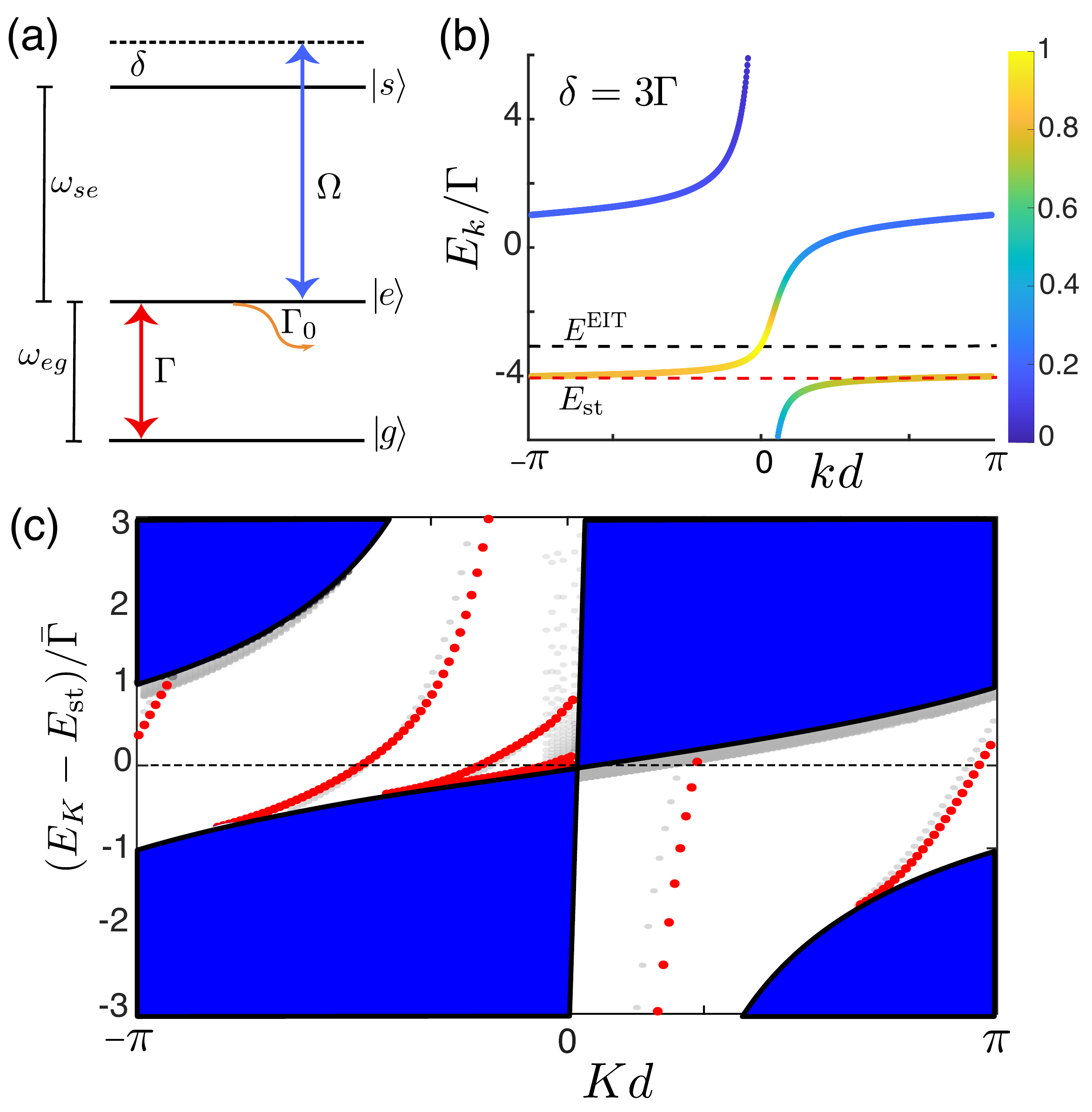}%
\caption{(a) Sketch of the three levels of a Rydberg atom involved in the scheme. (b) Single excitation dispersion relation associated to the Hamiltonian~\eqref{Haryd} for  $\Omega=2\Gamma$ and  $\delta=3\Gamma$. The color scale is associated to the fraction of population in the Rydberg state $|s\rangle$. The horizontal black dashed line indicates the EIT ($E^{\rm EIT}$) resonance while the red dashed line the Stark shifted one $E_{\rm st}$. (c) Two-excitation spectrum $E_K$~(relative to the Stark shifted resonance $E_{\rm st}$) of the effective spin model \eqref{heffr_off}, as a function of the center of mass momentum $K$ for $C_6/(d)^6=600\bar\Gamma$. In grey we have superimposed the equivalent spectrum (continuum and bound states) obtained by solving the full model with $\delta=-10\Gamma$ and $\Omega=\Gamma$ (see App.~\ref{App.A}). The black lines indicate the boundaries of the continuum states, obtained from the single excitation dispersion relation.}
\label{Fig.Rtran}
\end{figure}

It is instructive to first investigate the single excitation spectrum, in the case of a far off-resonant control field $|\delta|\gg \Omega$. The dispersion relation is derived in App.\ref{App.A} and plotted in Fig.~\ref{Fig.Rtran}(b) for a representative set of parameters.  It presents the usual three polariton branches \cite{lukinEIT,bienas}  and it  is characterized by the occurrence of a band that exhibits electromagnetically induced transparency~(EIT), which is centered around the EIT resonance condition $E^{\rm EIT}=-\delta$~\cite{lukinEIT} (black dashed line). 
Around this resonance, the polariton consists mostly of a Rydberg excitation, as indicated by the color scale of Fig.~\ref{Fig.Rtran}(b) associated to the amount of population on the $|s\rangle$ state. The excited state population $|e\rangle$ and its corresponding emission is suppressed via interference in excitation pathways, allowing this ``dark state'' polariton to propagate without loss~\cite{lukinEIT}.

Two or more dark state polaritons in a close vicinity of each other then strongly interact via the Rydberg interaction $V$ in~\eqref{Haryd}. This significantly affects the transparency condition and results in a strong photon-photon interaction, which can be either dissipative~\cite{Peyr}, for $|\delta |\ll\Gamma_0$, or dispersive in nature~\cite{first}, for $|\delta |\gg\Gamma_0$, depending on the detuning of the control field.
Quantum nonlinear optics has been extensively explored around the EIT transition. Notably, in the dispersive regime, this includes the observation of signatures of bound states, via photon bunching correlations in the outgoing field given a cw input~\cite{first,LiangBS}. Theoretically, these states, which can be of rich and varying nature, have been characterized via effective continuum theories~\cite{bienas,Efimov,magrebi}.\\

Here, in order to distinguish from previous work and to also explore connections with SIT, we will consider the nonlinear effects that arise around the narrow, effective two-level (Stark shifted) resonance, $E_{\rm st}\sim -\delta-\frac{\Omega^2}{\delta}$, indicated by the red dashed line of Fig.~\ref{Fig.Rtran}(b). Around this resonance, the excited state $|e\rangle$ can be effectively eliminated, and $|s\rangle$ can be considered the new, effective ``excited state'' of a two-level atom, with renormalized properties. This is evidenced by the similarity between the single-excitation dispersion relation in the vicinity of $E_{\rm st}$~(Fig.~\ref{Fig.Rtran}(b)), and that of Fig.~\ref{fig:setup_dis1ex}(b) for a chain of TLA coupled to a chiral waveguide. This intuition can be made more concrete by formally eliminating the far-detuned excited state, obtaining the effective Hamiltonian
\begin{equation}\label{heffr_off}
\begin{split}
\hat H_{\rm eff}=&\left(E_{\rm st}-i\frac{\bar\Gamma_0+\bar\Gamma}{2}\right)\sum_n\hat\sigma^n_{sg}\hat\sigma^n_{gs}-i\bar\Gamma\sum_{n>m}\hat\sigma^n_{sg}\hat\sigma^m_{gs}\\
&+\sum_{n<m}V(|x_n-x_m|)\hat\sigma^n_{sg}\hat\sigma^n_{gs}\hat\sigma^m_{sg}\hat\sigma^m_{gs}.
\end{split}
\end{equation}
This is indeed equivalent to the one given in Eq.~(\ref{Heffchi}) with renormalized emission rates $\bar\Gamma=\Gamma\Omega^2/\delta^2$ and $\bar\Gamma_0=\Gamma_0\Omega^2/\delta^2$.
The main differences with respect to Eq.~(\ref{Heffchi}) are the additional spin-spin Rydberg interaction term, and the~(large) additional independent emission $\bar\Gamma_0$. 

For more than one excitation, the Rydberg interaction allows for multiple bound state modes. This is illustrated in Fig.~\ref{Fig.Rtran}(c), where we plot the real part of the energy $E_K$ vs $K$ around the Stark shifted resonance $E_{\rm st}$.
A more in-depth discussion of these Rydberg bound states is provided in  App.~\ref{App.A} and App.~\ref{App.manybodyBSRyd}. However, the large additional dissipation $\Gamma_0$ acting on the bound states would make their observation challenging in free-space ensembles~(although perhaps Rydberg-like interactions could be emulated within waveguide QED~\cite{Efioptica,PRXcaneva} where the ratio of $\Gamma/\Gamma_0$ can be extremely high). Instead, in the following we will investigate the many-body limit, where it is known that SIT is robust in normal TLA ensembles with large dissipation, and examine how SIT changes in the presence of the long-range Rydberg interactions.

\subsection{Many-body dynamics}\label{Sec_many_rydberg}


\begin{figure}[h]
\includegraphics[width=0.48\textwidth]{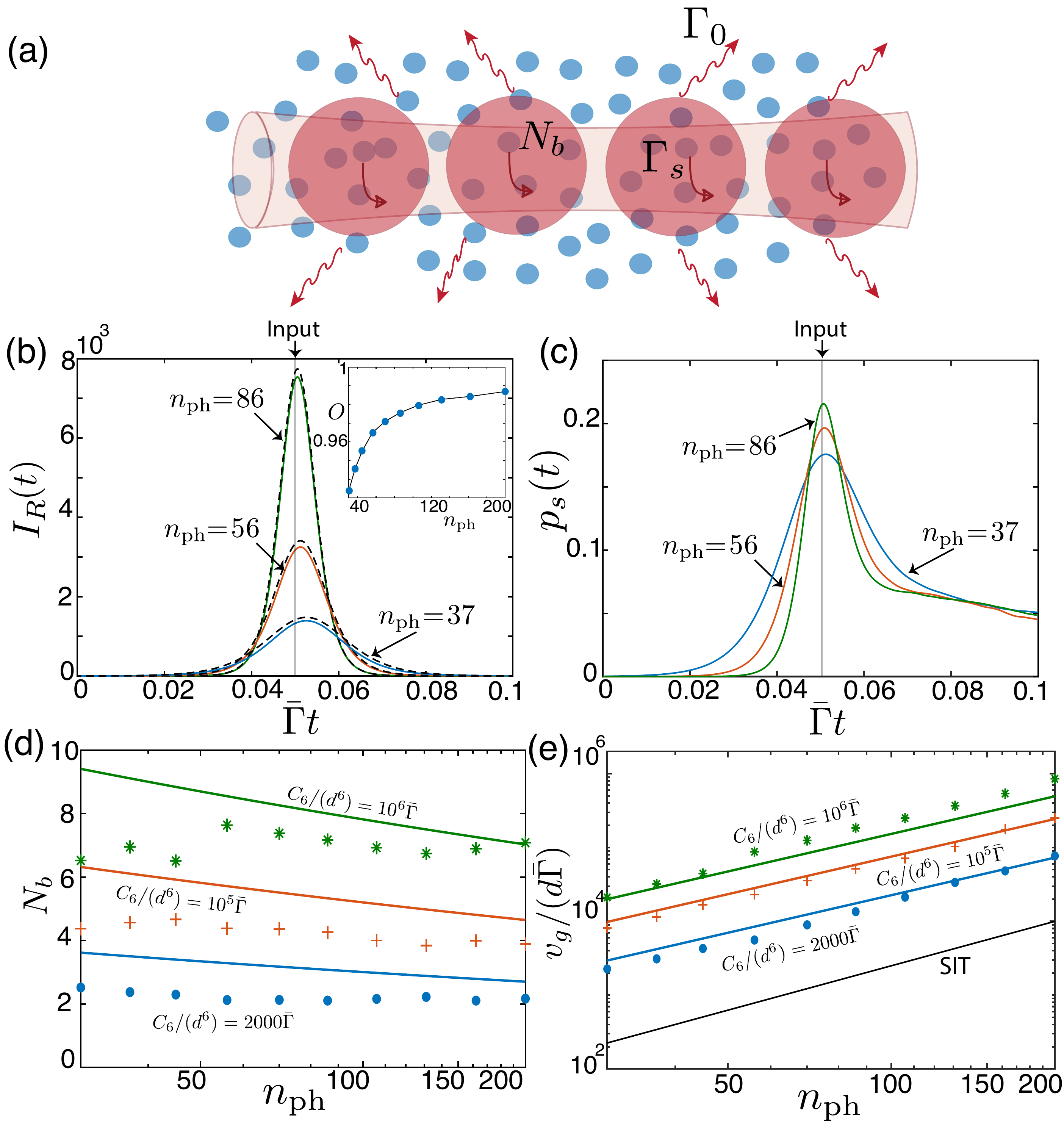}%
\caption{(a) Sketch of the effective array of superatoms arising in the Rydberg medium. (b)-(c) Output intensity (b) and  atomic population of an atom in the middle of the ensemble (c) as a function of time, for pulses of different average photon number. The form of the input field is given in Eq.~\eqref{ansatz_Rydberg},  with the soliton parameter $\alpha$ determined by a variational approach described in the main text. The input pulse peak is centered at $\bar\Gamma t_0=0.05$. In both plots we have added an external spontaneous emission rate of $\bar\Gamma_0=5\bar\Gamma$ and we set $C_6/d^6=10^5\bar\Gamma$. The continuous lines represents the results obtained numerically while the black dashed lines represent instead the delayed solitonic ansatz. In the inset of (b) we plot the overlap between the numerical obtained output and the solitonic ansatz as a function of  number of photons in the pulse.  (d)-(e) Estimated number of blockade atoms (d)  and  group velocity of the pulse (e) as a function of photon number for different Rydberg interaction strengths. The points are inferred from numerics, while the continuous lines in (d) and~(e) are obtained from our effective model described in the main text. In (e) the black line represents the standard SIT prediction for two-level atoms without Rydberg interactions. 
In all plots we take an array of $N=30$ atoms and we performed an MPS simulation of the full master equation by setting $D_{\rm max}=70$. 
In all plots regarding the soliton characterization, i.e. the inset of (b), and subfigures (d) and (e), we set $\bar\Gamma_0=0$ consistent with the standard SIT analysis.}
\label{Fig.tow_mf}
\end{figure}

Finding an analytic solution of the mean field equations~\eqref{eqMF} generalized to Rydberg interactions is more involved than in standard SIT due to the interactions. An approximate solution for these equations has been derived in Ref.~\cite{Rydberg-SIT} using a local field approximation that neglects two-body correlations. However, such a mean-field approach fundamentally breaks down in the case where $V$ is sufficiently large that a blockade radius emerges, a correlation effect in which two $|s\rangle$ excitations cannot be excited within a distance $r_b$ of one another due to the prohibitive energy cost. In particular, given a Rydberg excitation in the middle of the ensemble and a peak Rabi frequency $\Omega_r$, associated with the incoming pulse, the condition $V(r_b)=\Omega_r$ sets the distance at which the pulse can overcome the dispersive Rydberg energy shift and efficiently drive a second Rydberg excitation. The resulting ``blockade radius'' is then  $r_b=\left(C_6/\Omega_r\right)^{1/6}$~\cite{Lukin_Ryd}.
The blockade region consisting of $N_b=2r_b$ atoms acts like an effective, single two-level ``super-atom''~\cite{Hofferberth,Hofferberth2}, within which only a single collective Rydberg excitation can be generated. Furthermore, this excitation experiences a collectively enhanced emission rate $\Gamma_{\rm s}=N_b\bar{\Gamma}$ into the probe mode, while the dissipation $\bar{\Gamma}_0$ into other modes remains fixed.  For an extended ensemble the whole Rydberg medium can then be expected to consist of an array of $N/N_b$ effective super-atoms, as pictorially shown in Fig.~\ref{Fig.tow_mf}(a).
If this picture is correct, we might expect that a generalized version of SIT should emerge by simply re-scaling the decay rate in the  SIT solution of Eq.\eqref{eqSIT_solution}, 
where the standard SIT time delay in Eq.~\eqref{delaySIT} is modified to become $\tau=\frac{N}{N_b}\frac{4}{\bar\Gamma n^2_{\rm ph}N_b}$. This equation can be recast in terms of a more experimentally relevant quantity, the optical depth per blockade radius $D_b=2N_b\bar{\Gamma}/\bar{\Gamma}_0$~\cite{MPSJames,james_ryd}, such that
\begin{equation}\label{delayRydbergSIT}
 \bar\Gamma_0\tau=\frac{N}{N_b}\frac{8}{D_b n^2_{\rm ph}}.
\end{equation}
Note that $D_b$, $N/N_b$, the photon number in the pulse, and the free-space spontaneous emission rate $\bar{\Gamma}_0$ of the effective two-level transition all have well-defined meaning in an experimental setup of a Rydberg ensemble, which thus provides a connection between our microscopic spin model and a physical system.

In spite of the simplicity and semi-classical appearance of this guess, the blockade generates entanglement, and verifying this behavior requires a true many-body calculation, which we carry out using an MPS simulation with bond dimension $D\gg 1$ to capture the full correlated dynamics. The main steps of this procedure are presented in the following.

First we observe that, if the system supports a soliton-like solution, there should exist some input field, $\mathcal{E}_{\rm in}(t)$ (to be determined), which would result in an undistorted output, $I_{\rm sol}(t)=|\mathcal{E}_{\rm in}(t-\tau)|^2$, simply delayed by a time $\tau$ related to the group velocity of the pulse, $\tau=N/v_g$. As is it not feasible to numerically explore the infinite space of all input functions, here, we will restrict ourselves to a variational class, characterized by the parameter $\alpha$, which represents a generalization of the SIT solution
 \begin{equation}\label{ansatz_Rydberg}
\mathcal{E}_{\rm in}(t)=\frac{n_{\rm ph}\alpha\sqrt{\bar\Gamma}}{2}\operatorname{sech}\left(\frac{\bar\Gamma}{2}\alpha^2n_{\rm ph}(t-t_0)\right).
\end{equation}
This ansatz  recovers for $\alpha=1$ the usual SIT solution of Eq. \eqref{eqSIT_solution}, while for $\alpha=\sqrt{N_b}$ corresponds to our super-atom based hypothesis presented above.
Note that when acting on a single atom, this pulse violates the usual area law for the integrated Rabi frequency:
\begin{equation}
2\sqrt{\bar \Gamma}\int dt \mathcal{E}_{\rm in}(t)=\frac{2\pi}{\alpha}.\label{eq:SITarealaw}
\end{equation}
The ``best fit'' parameter $\alpha$ will in general be a function of the number of photons and the Rydberg interaction, i.e. $\alpha:=\alpha(n_{\rm ph},C_6)$. We determine the optimal value $\alpha_{\textrm{opt}}$  with a variational approach   that maximizes the overlap  $O=\int dt \,{\rm min}\{I_R(t),I_{\rm sol}(t)\}/\sqrt{\int dt I_R(t)\cdot\int dt I_{\rm sol}(t)}$ between the numeric and the expected output field intensity, $I_R(t)$ and  $I_{\rm sol}(t)$ respectively, where the pulse delay $\tau$ is numerically determined based on the position of the output intensity peak.
The inset of Fig.~\ref{Fig.tow_mf}(b) shows, for a chain of $N=30$ atoms, a progressively improving agreement between the output intensity and the expected soliton solution, when the number of photons is increased, with the optimal overlap reaching $O(\alpha_{\rm opt})\sim 99\%$ for $n_{\rm ph}=200$.
For the determined optimal values  $\alpha_{\textrm{opt}}$, we plot in Fig.~\ref{Fig.tow_mf}(b)  the corresponding output intensity at different pulse strengths. We also compare the numerically obtained output intensity with the expected damped solitonic solution $I_{\rm sol}(t)e^{-\bar \Gamma_0 \tau}$~(black dashed line). This approximately accounts for the external dissipation $\bar{\Gamma}_0=5\bar{\Gamma}$ introduced in the numerical simulations, and the two sets of curves are observed to agree well with one another. 
The difference compared to usual SIT is made evident in Fig.~\ref{Fig.tow_mf}(c), where we plot the time-dependent atomic excited population, $p_s(t)$, of an atom in the middle of the chain. We observe that, even in regimes of strong pulses, the excited population does not go to unity and back to zero as expected of a $2\pi$ pulse and discussed in Sec.\ref{Sec_TLA_many}, confirming the violation of the area law. 

In order to test the accuracy of the effective description based on the super-atom array, we compare its predictions with the numerical results describing the full correlated dynamics.
For a peak Rabi frequency $\Omega_r=n_{\rm ph}\alpha_{\textrm{opt}}\bar\Gamma$ associated with the pulse \eqref{ansatz_Rydberg}, the effective theory predicts an expected number of atoms within a blockade radius given by the solution of $r_b=N_b/2=\left(C_6/n_{\rm ph}\bar{\Gamma}\sqrt{N_b}\right)^{1/6}$. This is plotted  in Fig.~\ref{Fig.tow_mf}(d) (solid lines) as a function of the photon number $n_{\rm ph}$ for several different values of $C_6$, as indicated by different colors. The numerical estimate for $N_b$ can instead be inferred by 
the variationally determined value for $\alpha_{\textrm{opt}}$, using $N_b=\alpha_{\textrm{opt}}^2$, and is plotted as dots in  Fig.~\ref{Fig.tow_mf}(d).
 It is seen that the two approaches agree well with each other. Likewise, in Fig.~\ref{Fig.tow_mf}(e), we plot with points the group velocity, as given from the numerically determined delay by $v_g=N/\tau$. We then plot the same quantity $v_g=N/\tau$ in solid lines, where $\tau$ is given by the effective description~(\ref{delayRydbergSIT}), and where the expected $N_b$ is taken from the solid lines of Fig.~\ref{Fig.tow_mf}(d). Again, good agreement is seen. Finally, for comparison, we plot the expected group velocity for normal SIT~(without Rydberg interactions) in black, which is seen to be slower than in the blockaded case. These results shown that an effective description based on an array of Rydberg superatoms is able to capture the main emerging solitonic features of the propagating pulse.

We believe that the discrepancies between the numerical results and the effective description in Fig.~\ref{Fig.tow_mf}(d)-(e) can be mainly attributed to three sources of errors. The first is the intrinsic discreteness of our numerical model. For example, the moderate number of atoms $N_b\lesssim 10$ within a blockade region in our simulations suggests that moving to a continuum description should not be entirely accurate. A second error comes from the numerical imperfections in the evaluation of the optimal value $\alpha_{\textrm{opt}}$. Specifically, due to the large simulation complexity, the optimal value was obtained using a limited number of sampling points ($\sim 20$) for $\alpha$.  An additional~(non-numerical) discrepancy is that, similar to the case of SIT with normal TLA, the approach from quantum many-photon bound states to semi-classical SIT is a gradual one as a function of increasing photon number. For the lower range of photon numbers used in our simulations, it is possible that additional quantum many-body features are present, which would be interesting to pinpoint and explore further in future work.

We now consider the experimental feasibility of observing this many-body dynamics in a Rydberg ensemble. State-of-the-art experiments~\cite{LiangBS,Hofferberth,Hofferberth2} allow for a large optical depth per blockade radius, $D_b\gtrsim 1$, and multiple blockade regions in the ensemble, $N/N_b\sim 10$. For moderate photon numbers $n_{\rm ph}\sim 10^2$ in the pulse, it is then possible to acquire the time delay before full absorption takes place, $\bar\Gamma_0\tau\ll 1$. Separately, from Eqs.~(\ref{ansatz_Rydberg}) and~(\ref{delayRydbergSIT}), the ratio of the delay to the temporal width of the input pulse is given by $\tau/t_{\rm in}\sim (2/n_{\rm ph})(N/N_b)$. For parameters like above, this allows for the delay to be a reasonable, detectable fraction of the input pulse width.
Finally, as pointed out in Ref.~\cite{Rydberg-SIT}, this Rydberg SIT dynamics could also be probed without making use of an intermediate $|e\rangle$ state, by directly driving a ground-Rydberg state transition.

\section{Conclusions}\label{sec:conclusions}

In summary, we have presented a unified method to treat photon bound states within an atomic nonlinear medium, based upon a spin model formulation. This description  allows one to promptly identify the emerging correlated states in different  scenarios, such as an array of quantum emitters coupled to an optical waveguide and an ensemble of Rydberg atoms. In both cases, the formalism allows one to obtain the two-excitation bound state dispersion relation exploiting a convenient description in the relative coordinate frame, understand the effect of this dispersion relation on the correlation functions of outgoing fields, given an input pulse, and investigate the transition from few- to many-photon behavior. In the many-body case, we show how SIT or generalizations thereof emerge in all the systems studied. 

While the two-excitation limit and many-excitation, semi-classical SIT limit are possible to treat in semi-analytic fashion, in future work it would be interesting to develop techniques to better understand the intermediate case, where quantum effects in the many-photon pulse might still persist in the output. For example, it might be interesting to see whether the spin model might be more amenable to field theoretical techniques. Alternatively, it might be feasible to use time-independent MPS techniques to target and investigate the multi-excitation bound eigenstates themselves, in order to better understand their nature within the atomic medium.

\emph{Note added}. After the initial submission of this work a related preprint on photon bound states in waveguide QED systems  \cite{SorenwqedBS} reported similar results.

\section*{Acknowledgments}
The authors thank Sahand Mahmoodian, Anders S. S{\o}rensen, Angelo Carollo, Francesco Ciccarello  and Cosimo Rusconi for valuable discussions. G.C. acknowledge that results incorporated in this standard have received funding from the European Union Horizon 2020 research and innovation programme under the Marie Sklodowska-Curie grant agreement No. 882536 for the project QUANLUX.
D.E.C. acknowledges support from the European Union's Horizon 2020 research and innovation program, under FET-Open grant agreement No. 899275 (DAALI) and  European Research Council grant agreement No. 101002107 (NEWSPIN); the Government of Spain (Europa Excelencia program EUR2020-112155, Severo Ochoa program CEX2019-000910-S [MCIN/AEI/10.13039/501100011033], and MCIN Plan Nacional Grant PGC2018-096844-B-I00), Generalitat de Catalunya (CERCA program and AGAUR Project No. 2017-SGR-1334), Fundaci\'o Privada Cellex, Fundaci\'o Mir-Puig, and Secretaria d'Universitats i Recerca del Departament d'Empresa i Coneixement de la Generalitat de Catalunya, co-funded by the European Union Regional Development Fund within the ERDF Operational Program of Catalunya (project QuantumCat, ref. 001-P-001644).

\section*{Data  availability} The data presented in the figures of this manuscript are available on the link 
 DOI http://dx.doi.org/10.5281/zenodo.5771926.

\appendix

\section{Matrix product state simulation}\label{AppMPS}

\begin{figure*}[!t]
\includegraphics[width=0.9\textwidth]{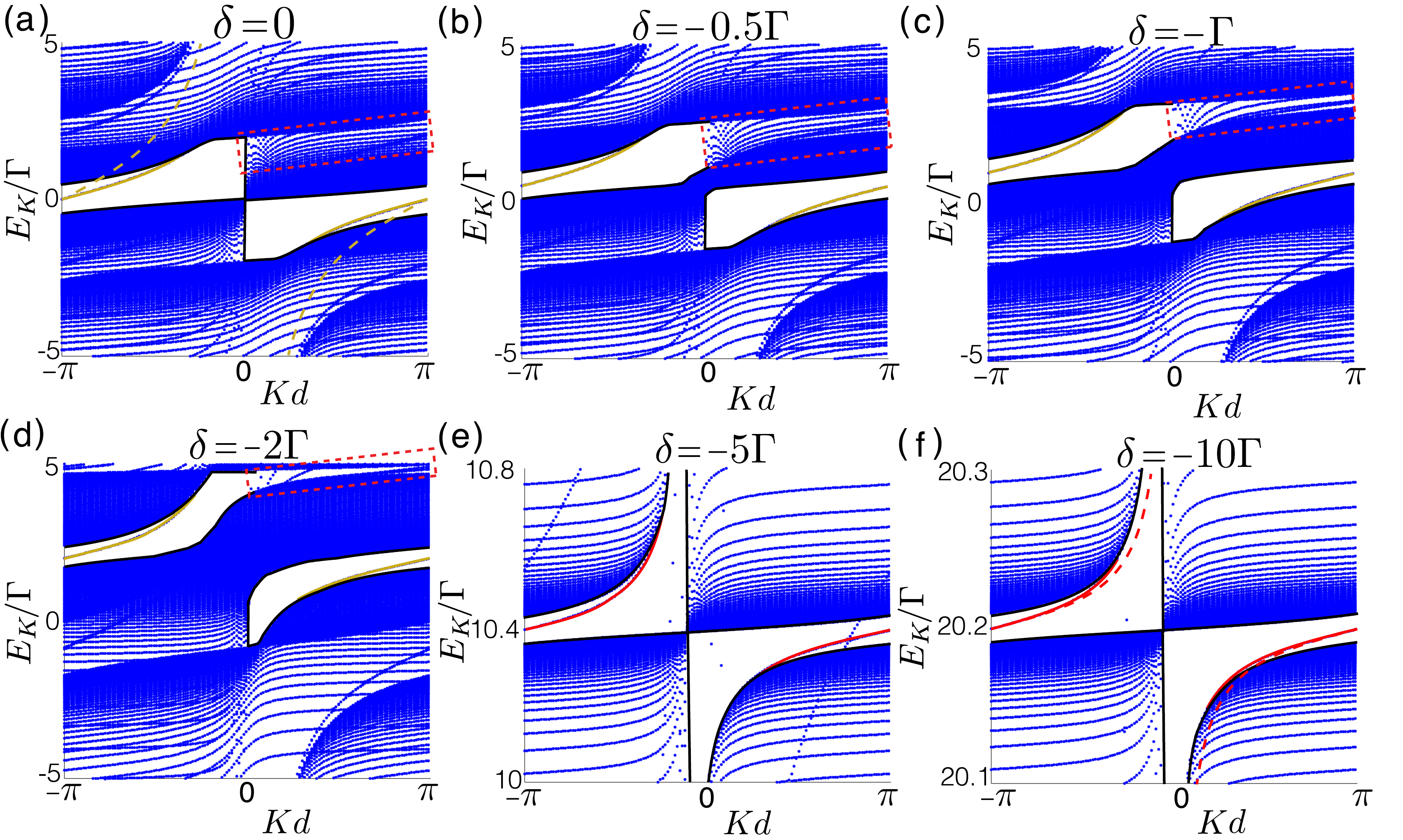}%
\caption{Spectrum $E_K$ of two-excitation eigenstates for the three-level atomic scheme presented in Sec.~\ref{Sec.modelRyd}, as a function of the center of mass momentum $K$ and for different detunings $\delta$ of the control field.  The continuous yellow line in (a)-(d) indicates the bound state obtained for the three level atom, which differs from the expected TLA bound state shown by the yellow dashed line in (a). In (e)-(f) the red continuous line indicates the bound state that emerges around the effective Stark-shifted TLA transition, as discussed in the main text. Its dispersion indeed approaches the one obtained with the rescaled effective model given in Eq.~\eqref{dis_EIT_BS} (red dashed line in (f)).
The dashed red lines in~(a)-(d) indicate the region where the bound states associated with the effective Stark-shifted TLA transition eventually arise for large $|\delta|$. In all plots we choose $C_6/(d)^6=0$ and $\Omega=\Gamma$.}
\label{Fig.Eit_band}
\end{figure*}

Solving the full system dynamics described by the master equation~\eqref{eq:MasterEq} for large atomic arrays, $N\gtrsim 20$, is extremely challenging with brute force numerical techniques.
To overcome this problem we adopt an MPS representation for the atomic array, which allows us to highly reduce the degrees of freedom needed to efficiently simulate the system. In particular, we make use of a quantum trajectories algorithm where the state of the system evolves under the effective Hamiltonian~\eqref{Heff}, along with stochastic quantum jumps as described in the details of Ref.~\cite{MPSJames}.
The main idea of the MPS representation consists in 
reshaping a generic quantum state $|\phi\rangle=\sum_{i_1,..i_N}\psi_{i_1,i_2,..i_N}|i_1,i_2,..i_N\rangle$ into a matrix product state of the form:
\begin{equation}
|\phi\rangle=\sum_{i_1,..i_N}A_{i_1}A_{i_2}...A_{i_N}|i_1,i_2,..i_N\rangle
\end{equation}
where, for each specific set of physical indices $\{i_1,i_2,..i_N\}$, the product of the $A_{ i_j}$ matrices gives back the state coefficient $\psi_{ i_1, i_2,.. i_N}$. Each matrix $A_{ i_j}$ has dimension $D_{j-1}\times D_{j}$ and finite-edge boundary conditions are assumed by imposing $D_{1}=1$ and $D_{N}=1$. 
The bond dimension $D_j$ is a crucial parameter because is directly connected to the entanglement entropy of the system. This implies that problems characterized  
by a limited entanglement entropy, as the ones considered in this paper, 
can be efficiently described by MPS ansatz with small bond dimension~\cite{Verstraete,Schollw}.
To compute the time evolution of the system we derive a matrix product operator (MPO) representation for the effective Hamiltonian and jump operators as described in Ref.~\cite{MPSJames}. The evolution of the initial ground state is then 
computed by using a Runge-Kutta method.

One difficulty for the simulation arises in the case of Rydberg atoms because the Rydberg interaction term, $\sum_{n<m}V(|x_n-x_m|)\hat\sigma^n_{sg}\hat\sigma^n_{gs}\hat\sigma^m_{sg}\hat\sigma^m_{gs}$ with  $V(|x_n-x_m|)=C_6/|x_n-x_m|$, does not have an exact MPO representation due to its power law nature. To overcome this issue we approximate the power law potential as a series of exponentials $V(x_r)\approx \sum_n\alpha_n\lambda_n^{x_r}$, as described in Refs.~\cite{Pirvu,Crosswhite,fro} and explicitly developed for atom-waveguide interactions in Ref.~\cite{james_ryd}. It is sufficient to truncate the maximum strength of the Rydberg interaction at the radius where the potential assumes a value an order of magnitude bigger than other characteristic scales, given by $\bar \Gamma$ and $\sqrt{N_b} n_{\rm ph}\bar \Gamma$ for the few and many body regimes respectively.

In addition to the Hamiltonian term, an MPO representation can be derived also for the output field and its associated correlation operators.
Both the time evolution and the computation of the observables at each time step are evaluated by applying an MPO to an MPS. This operation progressively increases the MPS bond dimension 
but, in order to keep the computation efficient, the bond dimension $D_j\leq D_{\rm max}$ can be truncated after each step.

In section~\ref{Sec_many_rydberg}, we employ a slightly different technique, by directly representing the density matrix in MPS form and solving the master equation without making use of the quantum trajectories algorithm, as explained in Ref.~\cite{james_ryd}. This choice, exploited for the plots of Fig.~\ref{Fig.tow_mf}, is well motivated by the relatively low maximum bond dimension required for the density matrix.

\section{Spin model approach for the Rydberg media}\label{App.A}
In this section we use the spin model to derive the single- and two-excitation spectrum for the three-level Rydberg atom scheme presented in Sec.~\ref{Sec.modelRyd}.

\subsection{Single excitation and EIT}
For a single excitation the Rydberg interaction term given in~\eqref{Haryd} does not play a role and it is convenient to rewrite the spin Hamiltonian~\eqref{Heff} 
in wavevector space $k$ where it reads:
\begin{equation}\label{Hrydk}
\hat H_{\rm eff}=\tilde J_k\sum_k\hat\sigma_{eg}^k\hat\sigma_{ge}^k -\delta\sum_k \hat\sigma^k_{se}\hat\sigma^k_{es}+\Omega\sum_k\left(\hat\sigma_{es}^k+\hat\sigma_{se}^k\right),
\end{equation}
where $\tilde J_k=-\frac{\Gamma}{2}\cot{(kd/2)}-i\Gamma_0/2$ coincides, except for the spontaneous emission term $\sim \Gamma_0$, with the dispersion for a chiral atomic array derived in Eq.~\eqref{dis_chi}. 
Hamiltonian \eqref{Hrydk} can be exactly diagonalized  by considering a linear combination of Rydberg and excited state excitations: $|\psi_k\rangle= \alpha_k|e_k\rangle+\beta_k|s_k\rangle$. This leads to the following coupled equations:
\begin{equation}
\begin{split}
&\left(E_k-\tilde J_k\right)\alpha_k=\Omega \beta_k\\
&\left( \delta+E_k\right)\beta_k=\Omega \alpha_k.
\end{split}
\end{equation}
These equations provide the  full dispersion divided in two contributions, which  we name upper~(U) and lower~(L):
\begin{equation}\label{disEITex}
E^{(U/L)}_k=\frac{\tilde J_k-\delta}{2}\pm\frac{1}{2}\sqrt{(\tilde J_k-\delta)^2+4(\Omega^2+\delta\tilde J_k)},
\end{equation}
associated to the  states: 
\begin{equation}\label{polariton}
|\psi^{(U/L)}_k\rangle=\Psi^{\dagger(U/L)}_k|0\rangle=\frac{(\delta+E^{(U/L)}_k)\hat \sigma^k_{ge} +\Omega \hat \sigma^k_{gr}}{\sqrt{|\delta+E^{(U/L)}_k|^2+\Omega^2}}|0\rangle.
\end{equation}
The combination of these two contributions  gives the dispersion relation  shown in Fig.~\ref{Fig.Rtran}(b), which  is characterized by three polariton branches, as usually obtained in literature \cite{lukinEIT,bienas} and by the occurrence of an EIT transmission band  centered at the EIT resonance $E^{\rm EIT}=-\delta$~ (dashed horizontal line). Note that similarly as discussed for the waveguide case, the polariton branches diverge at $k=0$, due to the Markov approximation, instead of following the light line with slope $c$. 
Around the EIT resonance the dispersion~\eqref{disEITex}  reduces to 
 \begin{equation}\label{disEIT}
E_k\simeq-\delta+\frac{\Omega^2}{\Gamma}kd+\frac{\Omega^2}{\Gamma^2}(\delta-i\Gamma_0/2) (kd)^2+...,
\end{equation}
and the state~\eqref{polariton} becomes mainly a Rydberg-like excitation 
with $\alpha_k\sim 0$.
The modes occurring around this resonance do not suffer from spontaneous emission into free space~($\Gamma_0$) due to the negligible excited state population, and these ``dark state'' polaritons allow the medium to be transparent to light in this frequency range~\cite{lukinEIT}. These dark state polaritons propagate in the medium with EIT group velocity $v_g=\partial E_k/\partial k|_{k=0}=\Omega^2d/\Gamma$ and effective mass $m_{\rm eff}=\hbar/(2\partial E^2_k/\partial k^2|_{k=0})=\hbar \Omega^2/(2 v_g^2\delta)$.
Note that  by solving the  full polaritonic model the EIT group velocity is known to be given by $v^{\rm full}_g=\frac{\Omega^2cd}{\Omega^2d+c\Gamma}$~\cite{lukinEIT}. This result reduces to the one  derived  with the spin model in the \qq{slow-light} limit, i.e. $\Omega^2/\Gamma\ll c/d$, which is equivalent to the Markov approximation.

\subsection{Two excitation spectrum}
The two-excitation subspace is spanned by the basis set $\{|e_ne_m\rangle,|s_ns_m\rangle,|e_n s_m\rangle,|e_m s_n\rangle\}$ with $n>m$. Similarly as done for the two-level atom array in Sec.~\ref{Secarray2ex},  we can re-parametrize the  eigenstates in the center of mass, $x_{\rm cm}=(x_m+x_n)/2$, and  relative coordinate, $x_r=|x_n-x_m|$, assuming a plane wave ansatz along $x_{\rm cm}$ of the form
\begin{equation}\label{ans_ryd}
\begin{split}
 &|\psi^{(2)}\rangle =\sum_{x_{\rm cm}}e^{iKx_{\rm cm}}\left(f_1(x_r)\hat\sigma^{(x_{\rm cm}-x_r/2)}_{eg}\hat\sigma^{(x_{\rm cm}+x_r/2)}_{eg}\right.\\
& \left. +f_2(x_r)\hat\sigma^{(x_{\rm cm}-x_r/2)}_{sg}\hat\sigma^{(x_{\rm cm}+x_r/2)}_{sg}\right.\\
&\left.+f_3(x_r)\hat\sigma^{(x_{\rm cm}-x_r/2)}_{eg}\hat\sigma^{(x_{\rm cm}+x_r/2)}_{sg}\right.\\
& \left. f_4(x_r)\hat\sigma^{(x_{\rm cm}-x_r/2)}_{sg}\hat\sigma^{(x_{\rm cm}+x_r/2)}_{eg} \right)|0\rangle,
\end{split}
\end{equation}
where $f_l(x_r)$ are generic functions of the relative coordinate. 
The effective single-excitation problem in the relative coordinate is more complicated than the case of TLA, as the excitation can occupy one of four distinct sectors characterized by the creation operators $\hat S^{\dagger}_{ee}$, $\hat S^{\dagger}_{ss}$, $\hat S^{\dagger}_{se}$ and $\hat S^{\dagger}_{es}$. With this notation the Hamiltonian for the relative coordinate reduces to
\begin{equation}\label{HKeff_ryd}
\begin{split}
&\hat H^K=-\sum_{r>0}\left[ \left(2\delta -V(x_r)\right) \hat S^{\dagger r}_{ss}\hat S^{r}_{ss}+ \delta \hat S^{\dagger r}_{se}\hat S^{r}_{se}+\delta \hat S^{\dagger r}_{es}\hat S^{r}_{es}\right]\\
&-i\frac{\Gamma}{2}\sum_{r,r'>0}\sum_{\epsilon=\pm1}\left[e^{-i\frac{K}{2}|x_r+\epsilon x_{r'}|}-e^{i\frac{K}{2}|x_r+\epsilon x_{r'}|}\right]\hat S^{\dagger r}_{ee}\hat S^{r'}_{ee}\\
&-i\frac{\Gamma}{2}\sum_{r>r'}\left[e^{-i\frac{K}{2}|x_r-x_{r'}|}\hat S^{\dagger r}_{es}\hat S^{r'}_{es}-e^{i\frac{K}{2}|x_r-x_{r'}|}\hat S^{\dagger r}_{se}\hat S^{r'}_{se}\right]\\
&-i\frac{\Gamma}{2}\sum_{r<r'}\left[e^{-i\frac{K}{2}|x_r-x_{r'}|}\hat S^{\dagger r}_{se}\hat S^{r'}_{se}-e^{i\frac{K}{2}|x_r-x_{r'}|}\hat S^{\dagger r}_{es}\hat S^{r'}_{es}\right]\\
&-i\frac{\Gamma}{2}\sum_{r,r'>0}\left[e^{-i\frac{K}{2}|x_r-x_{r'}|}\hat S^{\dagger r}_{es}\hat S^{r'}_{se}-e^{i\frac{K}{2}|x_r-x_{r'}|}\hat S^{\dagger r}_{se}\hat S^{r'}_{es}\right]\\
&+\Omega\sum_{r>0}\left[\hat S^{\dagger r}_{es}\hat S^{r'}_{ee}+\hat S^{\dagger r}_{se}\hat S^{r'}_{ee}+\hat S^{\dagger r}_{es}\hat S^{r'}_{ss}+\hat S^{\dagger r}_{se}\hat S^{r'}_{ss}+\rm H.c. \right],
\end{split}
\end{equation}
which depends parametrically on the center-of-mass momentum $K$. Eq.~\eqref{HKeff_ryd} can be diagonalized numerically.

\subsubsection{Two excitation spectrum for $C_6=0$}\label{App.RydEIT}

In Fig.~\ref{Fig.Rtran}(c) of the main text we presented the two-excitation spectrum of the Rydberg medium spectrum, obtained by diagonalizing the effective two-level model~\eqref{heffr_off}. Here we want to show how this spectrum arises, within the full three-level model, starting from the resonant case, $\delta=0$ and  progressively moving off resonance $\delta\gg \Gamma$. In Fig.~\ref{Fig.Eit_band} we plot the spectrum obtained by the diagonalization of Eq.~\eqref{HKeff_ryd} for the paradigmatic case where there is no Rydberg interaction, $C_6=0$.  In Fig.~\ref{Fig.Eit_band} (a) for $\delta=0$ we observe a bound state in the gap (continuous yellow line) with a dispersion that differs from the one of the TLA case (dashed yellow line) due to the three level nature of the medium. 
In the main text we discussed the off resonance regime where an effective (rescaled) TLA atom description can be used. The transition towards this regime is shown in  Fig.~\ref{Fig.Eit_band}(b)-(d), where the density of states inside the region highlighted by the red dashed lines decreases for increasing $\delta$.
Moving further off-resonance (panels (e)-(f)), we observe that a new gap arises in this region, which is captured by the effective TLA Hamiltonian~\eqref{heffr_off}.
This effective model, in absence of Rydberg interactions, owns a bound state solution (red dashed line in  Fig.~\ref{Fig.Eit_band} (f)) that follows the dispersion:
\begin{equation}\label{dis_EIT_BS}
\omega(K)=2\frac{\Omega^2}{\Delta_p}-2\Gamma\left(\frac{\Omega}{\Delta_p}\right)^2\cot(Kd/2)
\end{equation}
and well approximates the one obtained by the full diagonalization of  Eq.~\eqref{HKeff_ryd} (continuous red line).


 \begin{figure}
\includegraphics[width=0.48\textwidth]{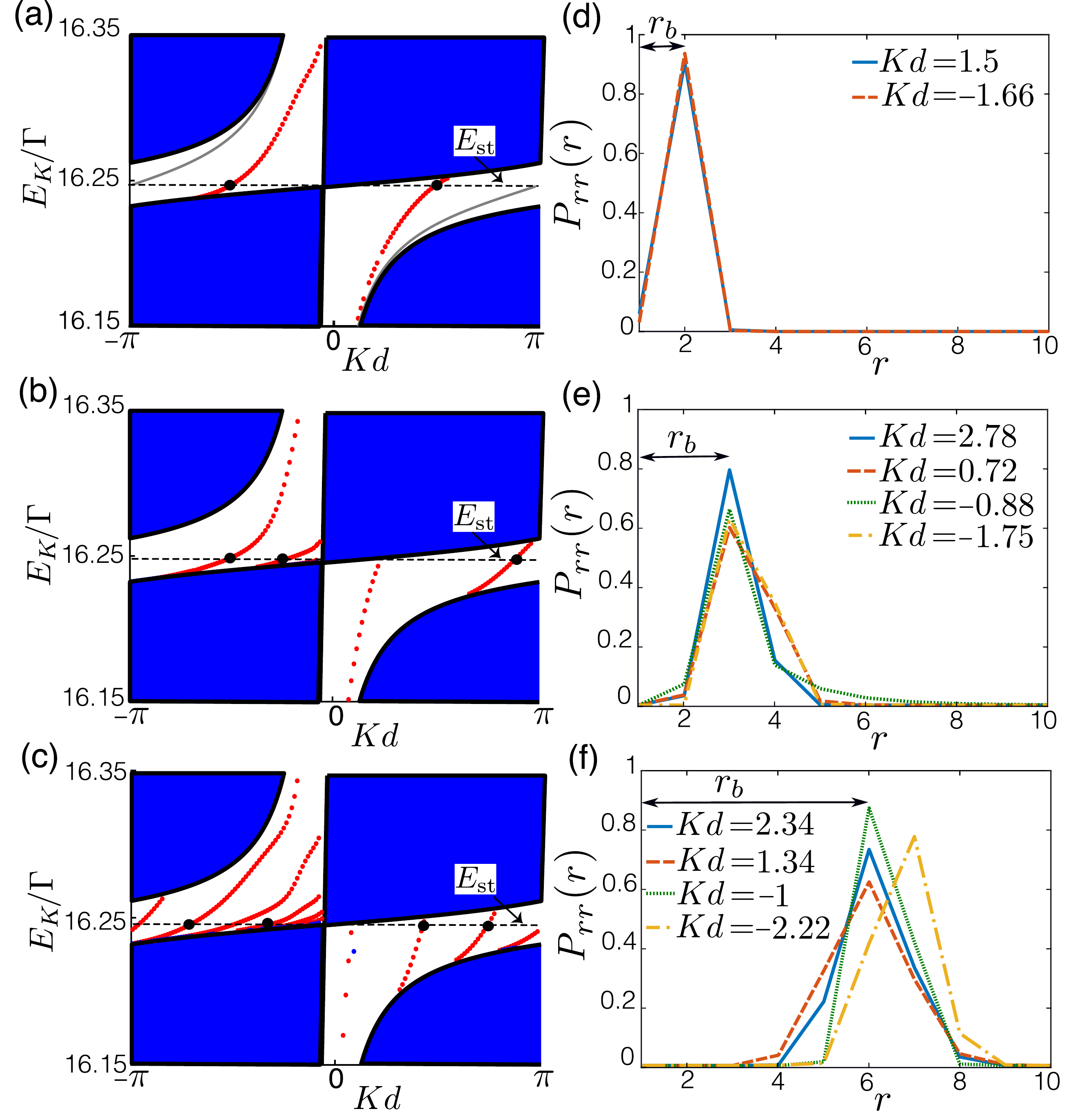}%
\caption{(a)-(c) Spectrum $E_K$ of two-excitation eigenstates versus center-of-mass momentum $K$ for $\delta=-8\Gamma$, $\Omega=\Gamma$ and different strengths of the Rydberg interaction: $C_6/d^6=0.1\Gamma$ in (a), $C_6/d^6=10\Gamma$ in (b) and $C_6/d^6=1000\Gamma$ in (c). Here, we focus on the region of the spectrum near the effective, Stark-shifted TLA transition. The grey line in (a) indicates the bound state dispersion relation in absence of Rydberg interaction due to the TLA nonlinearity, while the Rydberg interaction induced bound states are given in red. The black lines indicate the boundaries of the continuum states. (d)-(f) Rydberg population distribution as a function of the relative distance index $r=x_r/d$, for the corresponding strengths of the Rydberg interaction shown in (a)-(c). The bound states populations plotted in the panels correspond to specific values of $K$ and are highlighted in (a)-(c) by the black dots.}
\label{Fig.Rybderg_band_differentV}
\end{figure}

\subsubsection{Two excitation spectrum for $C_6\neq 0$}\label{App.RydEIT_loc}

We now investigate the spectrum of the two-excitation subspace in presence of Rydberg interactions. Different strengths of the Rydberg interaction affect the bound state dispersion relation as illustrated in Fig.~\ref{Fig.Rybderg_band_differentV}. In particular, in Fig.~\ref{Fig.Rybderg_band_differentV}(a), we see how a weak interaction is 
sufficient to drastically change the bound state dispersion~(red) compared to the one previously obtained for an array of two-level atoms~(grey). Increasing the interaction strength $C_6$, multiple bound states progressively arise with a dispersion that becomes steeper as they move further away from the continuum of states.
The increasing interaction strength also affects the population distribution in the relative coordinate. In particular, in Figs.~\ref{Fig.Rybderg_band_differentV}(d)-(f) one sees that the distribution $P_{rr}(r)$ of two Rydberg excitations exhibits a progressive growth in the separation between them. This localization distance is approximately given by the Rydberg blockade radius, $r_b=(C_6/\bar \Gamma)^{1/6}$, defined in the main text for the frequency regime close to the Stark-shifted resonance.

\section{Exciting multiple bound states in the Rydberg media}\label{App.manybodyBSRyd}
\begin{figure}
\includegraphics[width=0.48\textwidth]{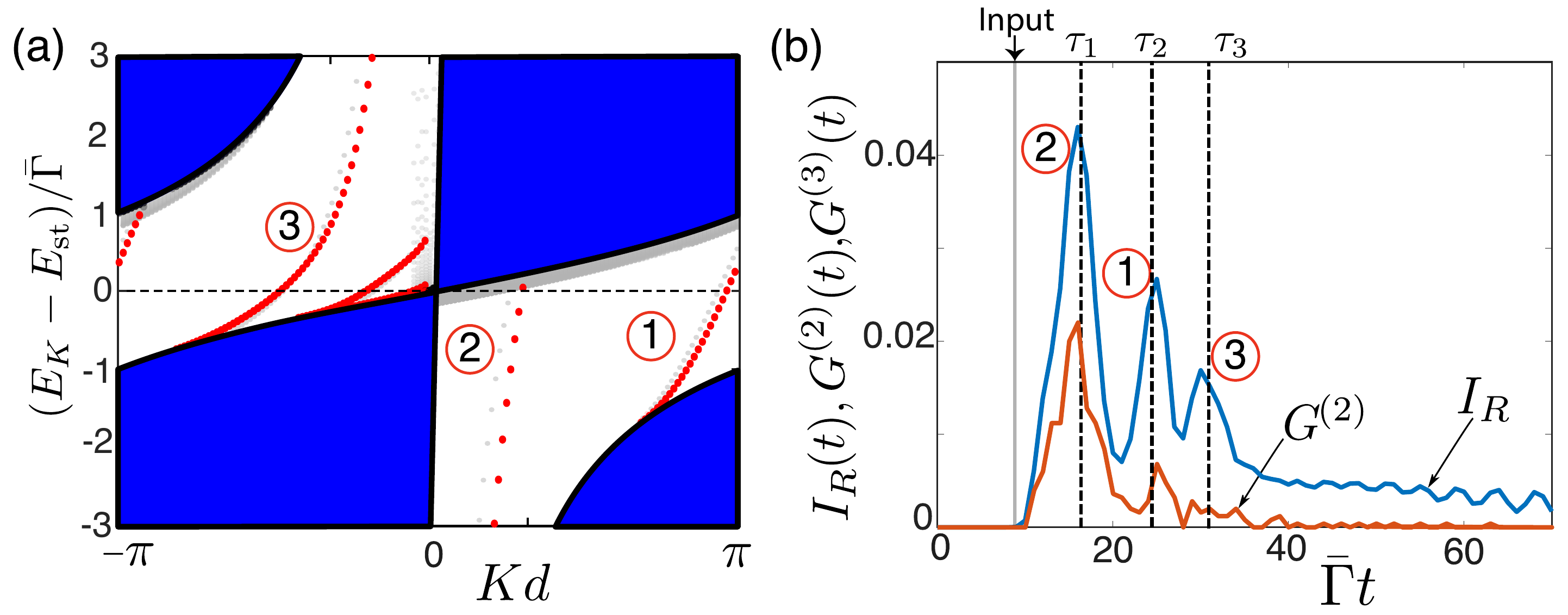}%
\caption{(a) Same eigenvalue spectrum as the one shown in Fig.~\ref{Fig.Rtran}(c) of the main text. (b) Output intensity  $I_R(t)$ and two-photon correlation function $G^{(2)}(t)$, due to a coherent Gaussian input pulse propagating through an array of $N=40$ atoms with $\sigma\bar\Gamma=2$ and $n_{ph}=1.0$. The expected time delays arising from excitation and propagation of bound states, $\tau_l=Nd/v^l_g(\omega_{in})$, are indicated by vertical dashed lines and are computed numerically from the group velocity of the bound state branches $l$ plotted in (a). The grey vertical line instead indicates the initial time of the input pulse set to  $\bar\Gamma t_0=10$. The simulation has been performed with an MPS based quantum trajectories algorithm involving $N_t=5000$ trajectories and maximum bond dimension $D_{\rm max}=40$.  }
\label{Fig.pulse_stark}
\end{figure}

Here, we show how the multiple two-excitation bound states occurring in the Rydberg system can be excited given an input pulse. To simulate the dynamics we make use of the effective TLA model~\eqref{heffr_off} and employ the MPS simulation as previously discussed.
The bound states shown in the spectrum of Fig.~\ref{Fig.Rtran}(c) (repeated in 
Fig.~\ref{Fig.pulse_stark}(a) for convenience) can be excited by sending a coherent Gaussian input pulse at the frequency $E_{\rm st}$ through the atomic array. In the ideal lossless regime, $\Gamma_0=0$, the time-dependent output intensity $I_R(t)$ and the equal time two-photon correlation function $G^{(2)}(t)$ exhibit isolated peaks that can be associated to the different bound states of Fig.~\ref{Fig.pulse_stark}(a) through their acquired time delay $\tau_l=Nd/v^l_g(\omega_{in})$, where $l$ is the label of the bound state branches (circled numbers in Fig.~\ref{Fig.pulse_stark}). In order to resolve the multiple bound states in the output field, we have considered in Fig.~\ref{Fig.pulse_stark}(b) a large array of $N=40$ atoms. This allows that the delay between two different peaks is bigger than the sum of their widths, a condition that can be recast in the form $Nd\gg r_b(v_g^l+v_g^m)/|v_g^l-v_g^m|$. 
 
In a large array, moderate values of external spontaneous emission $\Gamma_0/ \Gamma\sim 0.1$ can already damp and smear out the features in the output field. As discussed in Sec.~\ref{Sec_many_rydberg} in the main text, the effect of external dissipation decreases by going toward the many-body regime, where the stimulated emission rate into the waveguide, $\sim \Gamma n_{\rm ph}$, can overcome the free space emission for large photon number input pulses. On the other hand, this  few-excitation dynamics might be observable in circuit QED platforms where Rydberg-like interactions are implemented in arrays of artificial atoms coupled to a transmission line. It would be interesting to further investigate how different types of correlated light might be realized through the engineering of direct qubit-qubit interactions.

\end{document}